\documentstyle[12pt]{article}
\newcommand{\beq}{\begin{equation}}
\newcommand{\eeq}{\end{equation}}
\newcommand{\id}
 {i\kern.06em\hbox{\raise.25ex\hbox{$/$}\kern-.60em$\partial$}}
\newcommand{\as}{\!\not\!\! A}

\newcommand{\bs}{/\kern-.52em b}
\newcommand{\qs}{/\kern-.52em s}
\newcommand{\D}{{\cal{D}}}
\newcommand{\dv}{\!d^3\!x\,}
\newcommand{\Z}{{\cal Z}}
\renewcommand{\d}{\partial}
\newcommand{\dd}
{\kern.06em\hbox{\raise.25ex\hbox{$/$}\kern-.60em$\partial$}}

\newcommand{\J}{{\cal J}}
\newcommand{\tr}{\mathop{\rm tr}\nolimits}
\begin{document}
\title{ Bosonization in  $d>2$ dimensions \thanks{Talk delivered
at {\it Trends in Theoretical Physics - CERN - Santiago de Compostela -
La Plata Meeting, La Plata, April-May 1997}, to be published}}
\author{ Fidel A. Schaposnik
\thanks{Investigador CICBA, Argentina}
\\
{\normalsize\it
Departamento de F\'\i sica, Universidad Nacional de La Plata}\\
{\normalsize\it
C.C. 67, 1900 La Plata, Argentina}}
\date{}
\maketitle
\begin{abstract}
I discuss in this talk a bosonization approach
recently developed in refs.\cite{1}-\cite{7}. It
leads to the (exact) bosonization rule for fermion currents
in $d \geq 2$ dimensions and also provides a systematic
way of constructing the bosonic action in different regimes.
\end{abstract}
\newpage
\tableofcontents

\section{Introduction}

I describe in this talk recent work on bosonization
developed in colaboration with my friends {\bf Nino Brali\'c} from
Universidad Cat\'olica de Santiago de Chile, {\bf C\'esar Fosco}
 from Centro At\'omico de Bariloche, {\bf Eduardo Fradkin} from
the University of Illinois at Urbana Champaign,
 {\bf Jean Claude
Le Guillou} from LAPP-Annecy and ENS de Lyon,
{\bf Enrique Moreno} from
City Colege and Baruch College of the CUNY, and {\bf Virginia Man\'\i as}
and {\bf Carlos N\'u\~nez}
 from the University of La Plata. More details can be found
in references \cite{1}-\cite{7}.

{}~

Bosonization is a mapping of quantum field theories of fermion
fields onto equivalent theories of boson fields. Well-established
in $1+1$ dimensions, bosonization constitutes there one
of the main tools
available for the study of the non-perturbative behavior of
both quantum field theories \cite{coleman} and of condensed
matter systems \cite{mattis}. In dimensions other than
$1+1$, much less is known. In this talk I will precisely discuss
the issue of $d>2$ bosonization within the path-integral approach
making special emphasis in the $ d=3 $ case, of particular interest
in condensed matter problems. The bosonization approach that
I will present corresponds to the path-integral framework and
basically establishes an identity between the generating functional
of the fermionic theory and the generating functional of the
equivalent bosonic theory. From this relation, the recipe
for bosonization of fermion currents is derived and the
current commutator algebra is presented. To complete the
bosonization program one should also calculate the energy-momentum
tensor algebra. This is not done in this work.

Related and unrelated  approaches to $d>2$ bosonization can be
found in \cite{lut}-\cite{DNS5}.

\section{The method}

The method is straightforward. We shall basically consider
the  case of free fermions but we shall also discuss
an interacting (Thirring) model.

One starts from the fermion
Lagrangian for $N$ massive free fermions in $d$ dimensions,
\beq
L = \bar\psi (\id   + m ) \psi
\label{1}
\eeq
The corresponding generating functional reads
\beq
Z_{fer}[s] =  \!\!\int \!\!\D \bar \psi \D \psi
\exp[-\!\!\int_M d^dx \bar \psi (\id  + \qs + m ) \psi ]
\label{2}
\eeq
where $s_\mu$ is the source for fermion currents, $s_\mu = s_\mu^{a}t^a$
with $t^a$ the generators of a group $G$ and $M$ a $d$-dimensional
manifold.

Our derivation heavily relies on the invariance of the fermion measure
under \underline{local} gauge transformations $h(x) \in \hat{G}$
with $\hat G$ the group of continuous maps $M \to G$. This ensures that
\beq
Z[s^h] = Z[s]
\label{3}
\eeq
with
\beq
s^h_\mu = h^{-1} s_\mu h +  h^{-1} \partial_\mu h
\label{4}
\eeq
Evidently, fermions can be integated in (\ref{2}),
\beq
Z_{fer}[s] = \det  (\id  + \qs + m )
\label{5}
\eeq
and the determinant in the r.h.s. of eq.(\ref{5}) will be used
to introduce an auxiliary field $b_\mu$ taking
values in the Lie algebra of $G$ through the trivial formula
\beq
Z_{fer}[s] = \int  \D b_\mu \, \delta^{(d)} (b_\mu - s_{\mu})
\, \det (\id  + \bs + m )
\label{6}
\eeq
Now, it will be convenient to replace the delta function in (\ref{6})
as follows
\beq
 \delta^{(d)} (b_\mu - s_{\mu}) \to
\Delta[b] \, \delta^{(n)} \left(
\varepsilon_{{\mu_1} {\mu_2} \ldots {\mu_d}}
(f_{{\mu_1}{\mu_2}}[b] - f_{{\mu_1}{\mu_2}}[s] )
\right)
\label{7}
\eeq
Here we have used that the equation
\beq
f_{{\mu}{\nu}}[b]  = f_{{\mu}{\nu}}[s]
\label{8}
\eeq
has for $s_\mu \ne 0$ the unique solution
\beq
b_\mu = s_\mu
\label{9}
\eeq
and $ \Delta[b]$ is a Faddeev-Popov-like jacobian,
\beq
\Delta[b] = | \det (
 2 \varepsilon_{{\mu_1} {\mu_2} \ldots {\mu_d}} D_{\mu_1}[b]) |
\label{10}
\eeq
with $D_\mu[b]$ the covariant derivative,
\beq
D_\mu[b] = \partial_\mu + [b_\mu,~]
\label{11}
\eeq
We do not consider for the moment Gribov like problems that
could arise for certain manifolds and groups. Concerning
the delta function in the r.h.s. of eq.(\ref{7}), $n$
depends on the space-time dimensions according to
$ n = {d(d-1)}/2 $
since one needs for
enforcing eq.(\ref{8})
as many $\delta$-functions as independent components
the curvature has.

It is at this point that the bosonic field whose dynamics will be
equivalent to that of the original Fermi field comes into play. We
introduce it as a Lagrange multiplier $A_{\mu_{3} \ldots \mu_{d}}$
enforcing the $\delta$-function in the path-integral (\ref{6}),
\begin{eqnarray}
Z_{fer}[s] & = &
\int  \D b_\mu \D A_{\mu_{3} \ldots \mu_{d}} \, \det (\id  + \bs + m )
\Delta[b]
\nonumber\\
& & \exp \left[ \lambda \, tr  \!\int d^dx  A_{\mu_{3} \ldots \mu_{d}}
\varepsilon_{{\mu_1} {\mu_2} \ldots {\mu_d}}
(f_{{\mu_1}{\mu_2}}[b] - f_{{\mu_1}{\mu_2}}[s] ) \right]
\label{12}
\end{eqnarray}
Here $\lambda$ is a constant which can be adjusted so
as to obtain an adequate normalization for the currents.
Now, we rewrite eq.(\ref{12}) in the form
\beq
Z_{fer}[s]  =
\int  \D A_{\mu_{3} \ldots \mu_{d}} \, \exp(S_{bos}[A])
\exp \left(-  \lambda \, tr \!\int d^dx  A_{\mu_{3} \ldots \mu_{d}}
\varepsilon_{{\mu_1} {\mu_2} \ldots {\mu_d}}
f_{{\mu_1}{\mu_2}}[s]  \right)
\label{13}
\eeq
where the bosonic action is defined as
\begin{eqnarray}
\exp(S_{bos}[A]) & = & \int \D b_\mu  \det (\id  + \bs + m ) \Delta[b]
\nonumber\\
& &
\exp \left( \lambda \, tr \!\int d^dx  A_{\mu_{3} \ldots \mu_{d}}
\varepsilon_{{\mu_1} {\mu_2} \ldots {\mu_d}}
f_{{\mu_1}{\mu_2}}[b]  \right)
\label{14}
\end{eqnarray}
Formulae (\ref{13})-(\ref{14}) constitute our basic bosonization
recipe: eq.(\ref{13}) allows to compute fermion current correlation
functions in terms of the bosonic field $A$ and eq.(\ref{14})
gives the bosonic action defining the dynamics of $A$.
It can be now appreciated in what sense we consider our bosonization
recipe exact: we have arrived {\it with no approximation} to
a bosonization recipe of the form
\beq
\bar \psi \gamma_\mu t^a \psi \to
2 \, \lambda \, \varepsilon_{{\mu} {\mu_2} \ldots {\mu_d}}
\partial_{\mu_2} A^a_{\mu_{3} \ldots \mu_{d}}
\label{15}
\eeq
However, except in $d=2$ dimensions where we know how to compute
exactly the fermion determinant appearing in (\ref{14}) and to
resolve the path-integral defining $S_{bos}$, one should appeal
to some approximation scheme to evaluate the bosonic action
accompanying recipe (\ref{15}). This means that only
in $d=2$ dimensions the complete bosonization recipe
is exact.

It should be stressed that
the bosonization recipe (\ref{15})
should  be taken as illustrative of the bosonization since the
rigorous equivalence between the fermionic
and the bosonic theory is at the level of the generating
functional $Z_{fer}[s]$ of Green functions. It is from $Z_{fer}[s]$
written in the form (\ref{13}) that
one has to compute
current correlation functions  in the bosonic language.
Note also that  in writing
recipe (\ref{15}) we have ignored terms which are
non linear in the source.
Although correlation functions of currents acquire a
contribution from this terms, this contribution is irrelevant in the
calculation of conmutator
algebra since they have local support. This can be easily seen,
for example, using the BJL method (see \cite{6} for a discussion and
\cite{7} for  the application of the
BJL method within the present bosonization approach).
Concerning the bosonic field $A^a_{\mu_{3} \ldots \mu_{d}}$
note that  it corresponds to scalar fields in $d=2$ dimensions
(see ref.\cite{6} for details on how to make contact with
the usual bosonization rules), to a vector field in $d=3$
dimensions and to an antisymmetric
(Kalb-Rammond) field in $d > 3$ dimensions
\cite{3},\cite{5}.

\section{ The abelian case in $ d = 3 $ }
The Abelian case in $3$ dimensions is particularly simple. To begin with,
the bosonization recipe for the fermion current reads
\beq
\bar \psi \gamma_\mu \psi \to \pm \frac{i}{8\pi}
 \varepsilon_{\mu \nu \alpha} \partial_\nu A_\alpha
\label{16}
\eeq
where we have chosen $\lambda$ so as to make contact with the normalization
employed in ref.\cite{7}.
Concerning the bosonic action, $\Delta[b] $ is trivial so that eq.(\ref{14})
simply reads
\beq
\exp(S_{bos}[A]) = \int \D b_\mu  \det (\id  + \bs + m )
\exp \left( \mp \frac{i}{16 \pi} \, tr \!\int d^3x  A_\mu
\varepsilon_{\mu \nu \alpha}
f_{{\nu}{\alpha}}[b]  \right)
\label{17}
\eeq
or, calling
\beq
-\log\det (\id  + \bs + m )  = \int d^3x L[b]
\label{18}
\eeq
we can write
\beq
\exp(S_{bos}[A]) = \int \D b_\mu
\exp( - S_{eff}[b,A])
\label{19}
\eeq
where $S_{eff}$ is defined as
\beq
S_{eff}[b,A] =
\int d^3x ( L[b]
\pm \frac{i}{16 \pi} \, tr \, A_\mu
\varepsilon_{\mu \nu \alpha}
f_{{\nu}{\alpha}}[b]  )
\label{20}
\eeq
(The double sign in eqs.(\ref{17})-(\ref{20}) is included for
convenience, see the discussion below)

Being in general $L[b]$ non-quadratic in $b$ one cannot path-integrate
in (\ref{17}) so as to obtain $S_{bos}[A]$. We shall see however that
there is a change of variables allowing to decouple $A_\mu$ from $b_\mu$
in $S_{eff}[b,A]$ so that one can control the $A_\mu$ dependence
of $S_{bos}[A]$ without necessity of explicitly integrating over $b_\mu$.
Let us define a new variable $b'_\mu$ through the equation
\beq
b_\mu = (1-\theta) b'_\mu + \theta A_\mu + V_\mu[A]
\label{21}
\eeq
where $V_\mu[A]$ is a gauge invariant function of $A_\mu$ so that
$b'_\mu$, the variable which will replace $b_\mu$, transforms
as a a gauge field. $\theta$ is an arbitrary parameter to be adjusted
later. The idea is to choose $V_\mu$
so as to decouple  $b'_\mu$ from $A$. This amounts to
impose the following condition
\beq
\frac{\delta^2 S_{eff}}{\delta b'_\mu(x)\delta A_\nu(y)} = 0
\label{22}
\eeq
which in terms of $V_\mu$ reads
\begin{eqnarray}
& &  2i \lambda \varepsilon_{\rho \sigma
\alpha} \partial_x^\alpha \delta(x-y)
+ \theta _y \int d^3z \frac{\delta^2 L[b(z)]}{\delta
b_\rho(y)\delta b_\sigma(x)}
\nonumber \\
& & + \int d^3u \left( \int d^3z \frac{\delta^2 L[b(z)]}{\delta
b_\beta(u) \delta b_\sigma(x)} \right) \frac{\delta
V_\beta(u)}{\delta A_\rho(y)} = 0
\label{23}
\end{eqnarray}
To go on we need an explicit
and necessarily approximate expression for $L[b]$. If
we are interested in the large-distance regime of
the bosonic theory
we can use  the result first  \cite{Jac}-\cite{Red}
for the $d=3$ fermion
determinant as an expansion in inverse powers of the fermion
mass
\beq
L[b] = \mp \frac{i}{8\pi} \varepsilon_{\mu \nu \alpha }b_\mu
\partial _\nu b_\alpha + \frac{1}{24\pi |m|}f_{\mu \nu }^2[b]
+ O({1\over {m^2}})  \label{24}
\eeq
The first term in (\ref{24}) is the well-honored Chern-Simons
action introduced in \cite{Jac} as a way of generating
a mass for gauge fields in three dimensions. The
double sign in this term is originated in a regularisation ambiguity
characteristic of odd-dimensions (see ref.\cite{GRS}). The second one
corresponds to the leading parity-even contribution to the fermion
determinant.
One easily sees that if one
tries for $V_\mu[A]$
the functional form
\beq
V_\mu[A] = i \frac{C}{m} \varepsilon_{\mu \nu \alpha }
f_{\nu \alpha }[A]
\label{25}
\eeq
one gets, from the decoupling equation (\ref{23}),
\beq
C = \pm 1/3
\label{26}
\eeq
Then, if for simplicity one chooses $\theta = -1$,
the bosonic action for $A_\mu$ can be easily found to be
\beq
S_{bos}[A] = \pm \frac{i}{8 \pi} \int \dv \varepsilon_{\mu
\nu \alpha }A_\mu \partial _\nu A_\alpha  + \frac{1}{24 \pi |m|}
\int  \dv f_{\mu\nu}^2[A] + O(1/m^2)
\label{27}
\eeq
One
can in principle determine, following the same procedure, the
following terms in the $1/m$ expansion of $S_{bos}$ by including the
corresponding terms in the fermion determinant expansion.
This result extends that originally presented in ref.\cite{1}. It
shows that the bosonic counterpart of the
three dimensional free fermionic theory is,
to order $1/m^2$, a Maxwell-Chern-Simons theory which is equivalent,
as it is well-known, in turn to a self-dual system
\cite{VN}-\cite{DJ}.

Alternatively to the $1/m$ determinant expansion,
one can consider an expansion in powers of $b_\mu$ retaining
up to quadratic terms. The result can be written
in the form \cite{BFO}
\beq
L[b]= \frac{i}{2}\varepsilon _{\mu \nu \alpha }b_\mu
P\partial _\nu b_\alpha  + \frac{1}{  {4 |m|}}f_{\mu \nu
}[b]Qf_{\mu \nu }[b]
\label{28}
\eeq
where $P$ and $Q$ are functionals to be calculated within a loop
expansion,
\beq
P\equiv P\left( \frac{\partial ^2}{m^2}
\right)\;\;\;\;\;\;\;\;\;\;\;\;\;\;\;\;\;Q\equiv Q\left(
\frac{\partial^2}{m^2}
\right)
\label{29}
\eeq
Details of
the calculations of $P$
and $Q$  and results within the loop-expansion can be
found in refs. \cite{BFO},\cite{AFZ}.

In order to decouple the $b_\mu$ field one again proposes a change
of variables like in (\ref{21}) but now trying for $V_\mu $
the (gauge-invariant) functional form
\beq
V_\mu [A] =
\frac{i}{ m}\varepsilon _{\mu \nu \alpha }
R f_{\nu \alpha }[A]=2\frac{i}{m}\varepsilon _{\mu \nu \alpha }
R \partial _\nu A_\alpha
\label{30}
\eeq
with
\beq
R \equiv R\left( \frac{\partial ^2}{m^2} \right)
\label{31}
\eeq
One finds, from the decoupling conditions (\ref{23}),
\begin{eqnarray}
&&
\frac{\delta^2 S_{eff}[b',A]}{\delta A_\rho (y)\delta b'_\sigma(x)}
= \nonumber \\
& & (1-\theta )\;\left( \;i \varepsilon _{\rho \sigma \alpha}
\left(
2 \lambda + \theta P - 2 \frac{\partial ^2}{m^2} Q R
\right)
\right. \partial _\alpha \delta (x-y)
\nonumber
\\
& &\left. + \; \frac{2}{m}
\left( \frac{1}{2} \theta Q - P R  \right)
\left( {\partial _\rho \partial _\sigma
-\delta _{\rho \sigma }\partial ^2}\right)\delta (x-y)\;
\right)\;\;\;=0
\label{32}
\end{eqnarray}
$\theta$ being here a functional of $\partial ^2/m^2$.
The solution of this equation is
\beq
R=-\lambda \frac{Q}{\left( {P^2-{\frac{\partial ^2}{m^2}}Q^2}
\right)}\;\;\;\;\;\;\;\;\;\;\;\;\theta =-2\lambda
\frac{P}{\left(
{P^2-{\frac{\partial ^2}{m^2}}Q^2} \right)}
\label{33}
\eeq
With this choice, the change of variables decouples the $b_\mu$
integration so that one can finally get the bosonic action for
$A_\mu$ which now reads
\begin{eqnarray}
 S_{bos}[A] & = & \int \dv \left(
-(2 \lambda )^2\frac{i}{2}\varepsilon _{\mu \nu \alpha }
A_\mu \frac{P}{\left( {P^2-{\frac{\partial ^2}{m^2}}Q^2}
 \right)}\partial
_\nu A_\alpha  \right.
\nonumber \\
& & \left. + (2 \lambda )^2\frac{1}{4 m}
f_{\mu \nu }[A]
\frac{Q}{\left( P^2-\frac{\partial ^2}{m^2} Q^2 \right)}
f_{\mu \nu}[A] \right)
\label{34}
\end{eqnarray}
This result coincides with that found in ref.\cite{BFO},
obtained by a direct functional integration on $b_\mu$.
As it was proven in this last work, it corresponds
for massless fermions
to the bosonization action  proposed in
ref.\cite{Mar} since in the $ m \to 0$ limit eq.(\ref{34}) takes
the form
\beq
S_{bos} =   \frac{2}{\pi} \int d^3x (\frac{1}{4}
 F_{\mu \nu}
\frac{1}{\sqrt{-\partial^2}} F_{\mu \nu} -
\frac{i}{2} \epsilon_{\mu \nu \lambda}
A_\mu \partial_\nu A_\lambda)
\label{35}
\eeq

\section{Interacting models}
One can apply the bosonization approach described above to analyse
interacting fermionic models. Let us consider for example the
Thirring model with a current-current interaction Lagrangian
of the form
\beq
{L}_{int}=  -\frac{g^2}{2N}j_{\mu}j_{\mu}
\label{36}
\eeq
where $\psi^i$ are N two-component Dirac spinors and $j^{\mu}$ the $U(1)$
current,
\beq
j_{\mu}=\bar\psi^i\gamma^{\mu}\psi^i.
\label{37}
\eeq
The coupling constant $g^2$ has dimensions of inverse mass.
(Although non-renormalizable by power counting, four fermion
interaction models in $2+1$ dimensions are known to be
renormalizable in the $1/N$ expansion \cite{Gross}.)
One can directly apply the bosonization recipe found in
the precedent section to this interaction Lagrangian. Making
a choice of $\lambda$ so as to coincide with the
normalization in \cite{1}, this meaning
\beq
j_\mu \to i \sqrt{\frac{N}{4\pi}} \varepsilon_{\mu \nu \alpha}
\partial_\nu A_\alpha
\label{38}
\eeq
one has
\beq
{ L}_{int} \to   \frac{g^2}{16\pi } F^2_{\mu \nu}
\label{39}
\eeq
Alternatively, one can eliminate the quartic fermionic interaction by
introducing an auxiliary field $a_\mu$ via the identity
\beq
\exp(\int \frac{g^2}{2N}j^{\mu}j_{\mu} \dv)=\int\D a_{\mu}
\exp[-\int (\frac{1}{2}a^{\mu}a_{\mu}+\frac{g}{\sqrt{N}}j^{\mu}a_{\mu})\dv]
\label{40}
\eeq
and then proceed to integrate fermions as in the free  case
thus obtaining a determinant in which the $a_\mu$ field can be
eliminated by a shift $b_\mu \to b_\mu - a_\mu$. One confirms
in this way that the bosonization recipe (\ref{39}) is correct
so that the three dimensional Thirring model is
equivalent, in the $1/m$ approximation to a Maxwell-Chern-Simons model.
To leading order in $1/m$ we can then write (after rescaling the field
$A_\mu$)
\beq
L_{Th} \to \frac{1}{4} F_{\mu \nu}^2  \pm i
\frac{2\pi}{g^2} \epsilon^{\mu\alpha\nu}A_{\mu}\partial_{\alpha}A_{\nu}
\label{41}
\eeq
We can give now a first application of the bosonization formulas
and, in this way, explore their physical content. The Lagrangian
in (\ref{41}) has a Chern-Simons term which
controls its long distance behavior. It is well
known\cite{pol,witten} that the Chern-Simons gauge theory is a
theory of knot invariants which realizes the representations of
the Braid group. These knot invariants are given by expectation
values of Wilson loops in the Chern-Simons gauge theory. In
this way, it is found that the expectation values of the Wilson
loop operators imply the existence of excitations with
fractional statistics. Thus, it is natural to seek the
fermionic analogue of the Wilson loop operator $W_{\Gamma}$
which, in the Maxwell-Chern-Simons theory is given by
\beq
W_{\Gamma}=\langle \exp\{i {\frac{\sqrt N}{g}}\oint_{\Gamma}
A_\mu dx^\mu \}\rangle \label{42}
\eeq
where $\Gamma$ is the
union of a an arbitrary set of closed curves (loops) in three
dimensional euclidean space. Given a closed loop (or union of
closed loops ) $\Gamma$, it is always possible to define a set
of open surfaces $\Sigma$ whose boundary is $\Gamma$, {\it
i.e.} $\Gamma= \partial \Sigma$. Stokes' theorem implies that
\begin{eqnarray} \langle \exp \{ i {\frac{\sqrt N}{g}}
\oint_{\Gamma} A_\mu dx^\mu \}\rangle &=& \langle \exp \{ i
{\frac{\sqrt N}{g}} \int_{\Sigma} dS_\mu \epsilon^{\mu \nu
\lambda} \partial_\nu A_\lambda \}\rangle \nonumber \\
&=&\langle \exp \{i {\frac{\sqrt N}{g}} \int \dv \epsilon^{\mu
\nu \lambda} \partial_\nu A_\lambda  \; b_\lambda \}\rangle
\label{43}
\end{eqnarray}
is an identity. Here
$b_{\lambda}(x)$ is the vector field \beq b_\lambda(x)=
n_\lambda (x) \delta_{\Sigma}(x) \label{support} \eeq where
$n_\lambda$ is a field of unit vectors normal to the surface
$\Sigma$ and $\delta_\Sigma(x)$ is a delta function with
support on $\Sigma$. Using eq.(\ref{38}) we find that this
expectation value becomes, in the Thirring Model, equivalent to

\beq W_{\Gamma}=\langle \exp \{ i {\frac{\sqrt
N}{g}}\oint_{\partial \Sigma} dx_\mu A^\mu \}
\rangle_{MCS}=\langle \exp \{ \int_{\Sigma} dS_\mu {\bar \psi}
\gamma^\mu \psi \}\rangle_{Th}
\label{44}
\eeq
More
generally we find that the Thirring operator
${\cal W}_{\Sigma}$
\beq {\cal W}_{\Sigma}=\langle \exp
\{ q
\int_{\Sigma} dS_\mu {\bar \psi} \gamma^\mu \psi \}\rangle_{Th}
\label{thirringop} \eeq
obeys the identity
\beq \langle \exp \{
q \int_{\Sigma} dS_\mu {\bar \psi} \gamma^\mu \psi
\}\rangle_{Th}= \langle \exp \{ i q{\frac{\sqrt N}{g}}
\oint_{\partial\Sigma} A_\mu dx^\mu \}\rangle_{MCS}
\label{45}
\eeq
for an arbitrary fermionic charge $q$.

The identity (\ref{45}) relates the flux of the fermionic current
through an open surface $\Sigma$ with the Wilson loop operator
associated with the boundary $\Gamma$ of the surface.
The Wilson loop operator can be trivially calculated in the
Maxwell-Chern-Simons theory. For very large and smooth loops
the behavior of the Wilson loop operators is dominated by the
Chern-Simons term of the action. The result is a topological
invariant which depends only on the linking number $\nu_{\Gamma}$
of the set of curves $\Gamma$ \cite{pol,witten}.
By an explict calculation one finds
\beq
\langle \exp \{ q \int_{\Sigma} dS_\mu {\bar \psi} \gamma^\mu \psi \}
\rangle_{Th}=
\exp \{ \mp i \nu_{\Gamma} {\frac{Nq^2}{8 \pi}}\}
\label{46}
\eeq
This result implies that the non-local Thirring loop
operator ${\cal W}_{\Sigma}$ exhibits fractional statistics
with a statistical angle $\delta={{Nq^2}/{8 \pi}}$.
The topological significance of this result bears close resemblance
with the bosonization identity in $1+1$ dimensions between the
circulation of the fermionic current on a closed curve and the
topological charge (or instanton number) enclosed in the interior
of the curve \cite{colemanlibro}. From the point of view of the
Thirring model, this is a most surprising result which reveals
the  power of the bosonization identities. To the best of our
knowledge, this is the first example of a purely fermionic
operator, albeit non-local, which is directly related to a
topological invariant.

\section{Current algebra}

We have seen in precedent sections that in $3$ dimensional space-time
the fermion action bosonizes, in the large $m$ limit, to a
Maxwell-Chern-Simons theory. Now, the gauge invariant algebra of
such theory has been studied in refs.\cite{Jac},\cite{DJ}.
One has for instance with our conventions,
\beq
[E_i({\vec x},t),B({\vec y},t)] =
 - 3  \, \vert m \vert \epsilon_{ij}
 \partial_j\delta^{(2)}({\vec x} - {\vec y})
\label{47}
\eeq
If one now
relates the electric field $E_i = F_{i0}$ and the magnetic field
$B = \epsilon_{ij}\partial_iA_j$  to the fermionic
currents through the bosonization recipe for the fermion current,
\beq
j_o \to \frac{1}{\sqrt{4\pi}} B
\label{48}
\eeq
\beq
j_i \to \frac{1}{\sqrt{4\pi}}
 \epsilon_{ij}E_j
\label{49}
\eeq
then, the resulting fermion current commutator
algebra is not
the
one to be expected for three-dimensional free fermions. Indeed,
the $d=3$ fermion current algebra should contain an infinite
Schwinger term
\cite{BJ}-\cite{dVG} which is
absent in eq.(\ref{47}).
The point is that calculations
leading to a bosonic theory
of the Maxwell-Chern-Simons type
are valid only for large fermion mass while calculation of
equal-time current commutators imply, as we shall see,
a limiting procedure which cannot be naively followed
for large masses.

Since the exact bosonic partition function
is much too complicated to
handle, a possible strategy is to
use the quadratic (in auxiliary fields)
approximation mentioned in the
precedent section working with an
arbitrary (not necessarily large) mass so as to
obtain a bosonized version of the original fermionic
model in which the equal-time limit can be safely taken.
One should then compute current commutators for this bosonized theory,
and test whether they coincide with those satisfied
by fermionic currents in the original model. Details
of this calculation can be found in \cite{4}. I will just
sketch here the principal steps leading to the consistent equal-time
current commutators in the bosonic language.

As explained above, within the quadratic (in $b_\mu$) approximation,
one can write the fermionic partition function in terms of the
bosonic fields $A_\mu$ in the form \cite{BFO}
\begin{eqnarray}
Z_{fer} &=& \int DA_{\mu} \exp \left[-\int d^3x (\frac{1}{4}F_{\mu\nu}
C_{1}F_{\mu\nu} - \frac{i}{2}A_{\mu}C_{2}\epsilon_{\mu\nu\lambda}
\partial_{\nu}A_{\lambda}  \right. \nonumber\\
&  & \left. + i s_{\mu}
\epsilon_{\mu\nu\lambda}\partial_{\nu}A_{\lambda}) \right]
\label{50}
\end{eqnarray}
with $C_1$ and $C_2$ now given through their momentum-space
representation ${\tilde C}_1$ and ${\tilde C}_2$
\beq
{\tilde C}_{1}(k) = \frac{1}{4\pi} \,
\frac{{\tilde F}(k)}{k^{2} {\tilde F}^2(k) + {\tilde G}^2(k)}
\label{51}
\eeq
\beq
{\tilde C}_{2}(k) = \frac{1}{4\pi} \,
\frac{{\tilde G}(k)}{k^{2}{\tilde
F}^2(k) +
{\tilde G}^2(k)}
\label{52}
\eeq
and ${\tilde F}(k)$ and ${\tilde G}(k) $ given by \cite{BFO}
\[
{\tilde F} (k) \;=\; \frac{\mid m \mid}{4 \pi k^2} \,
\left[ 1
- \displaystyle{\frac{1 \,-\,\displaystyle{\frac{k^2}{4 m^2}}}{(
\displaystyle{\frac{k^2}{4 m^2}})^{\frac{1}{2}}}} \, \arcsin(1\,+
\, \frac{4 m^2}{k^2})^{-\frac{1}{2}} \right] \;,
\]
\[
{\tilde G} (k) \;=\; \frac{q}{4 \pi} \,+\, \frac{m}{2 \pi \mid k \mid}
\, \arcsin (1 \, + \, \frac{4 m^2}{k^2} )^{- \frac{1}{2}}
\]

Being quadratic in $A_\mu$,  eq.(\ref{50}) can be integrated
leading to
\beq
Z_{bos}[s] = \left[ det D_{\mu\nu}\right]^{- \frac{1}{2}}
\exp\left[\frac{1}{8\pi}
\int d^3x d^3y\partial_{\nu} s_{\mu}(x)\epsilon_{\mu\nu\lambda}
D_{\lambda\rho}^{-1}(x,y)\partial_{\sigma} s_{\tau}(y)
\epsilon_{\rho \sigma \tau}\right]
\label{53}
\eeq
where
$D_{\mu\nu}^{-1}$ is just the propagator of the bosonic action
which, in the Lorentz gauge we adopt from here on, reads
\beq
D_{\mu\nu}^{-1}(x,y) = \int \frac{d^{3}k}{{(2\pi})^3}
\left[P(k) g_{\mu\nu}  + Q(k) k_\mu k_\nu
+ R(k) \epsilon_{\mu\nu\alpha}k_{\alpha}
\right]\exp{i k(x - y)}
\label{54}
\eeq
with
\beq
P(k) = \frac{{\tilde C}_1(k)}{k^2 {\tilde C}_1^2(k) +
{\tilde C}_2^2(k)}
= 4\pi {\tilde F}(k)
\label{55}
\eeq
\beq
Q(k) = \frac{{\tilde C}_1(k)}{k^2 {\tilde C}_1^2(k) +
 {\tilde C}_2^2(k)}
\left(\frac{{\tilde C}_2(k)}{k^2{\tilde C}
_1(k)}\right)^2
\label{56}
\eeq
\beq
R(k) = \frac{{\tilde C}_2(k)}{k^2(k^2 {\tilde C}_1^2(k) +
{\tilde C}_2^2(k))}
\label{57}
\eeq
Let us
briefly recall how one can compute current commutators
within the path-integral scheme using the so-called
BJL method \cite{BJL}-\cite{Jac2}. To this end we define
the correlator
\beq
G_{\mu\nu}(x,y) = \left. \frac{\delta^{2} log Z_{fer}[s]}{\delta
 s_{\mu}(x)
\delta s_{\nu}(y)}\right|_{s=0}
\label{58}
\eeq
from which one can easily derive equal time current commutators
using the relation
\beq
<[j_0({\vec x},t),j_i({\vec y},t)]> =
\lim_{\epsilon \to 0^+}
[G_{0i}(\vec{x},t + \epsilon;{\vec y},t) -
G_{0i}(\vec{x}, t-\epsilon; {\vec y},t)]
\label{59}
\eeq
The current commutator evaluated using eqs.(\ref{58})-(\ref{59})
corresponds to  $Z_{fer}[s]$
written in terms of bosonic fields.
That is, eq.(\ref{59}) gives the
equal-time commutator for the bosonic currents
$j_\mu = (1/\sqrt{4\pi}) \epsilon_{\mu \nu \alpha} \partial _\nu
A_\alpha $. This result should
then be
compared with that arising in the original $3$-dimensional fermionic
model for which $j_\mu = - i\bar \psi \gamma_\mu \psi$ \cite{dVG}.

Starting from eqs.(\ref{53})-(\ref{54}) and using the
BJL method we get, after some calculations,
\beq
G_{\mu \nu}(x,y) =
-\frac{1}{4 \pi} \epsilon_{\mu \alpha \rho} \epsilon_{\nu \beta \sigma}
\partial_\alpha \partial_\beta D^{-1}_{\rho \sigma}
\label{60}
\eeq
or
\beq
G_{\mu \nu}(x,y) = \frac{1}{4 \pi}
\int \frac{d^3k}{(2\pi)^3} [P(k)(k^2 g_{\mu \nu} - k_\mu k_\nu)
+k^2 R(k) \epsilon_{\mu \nu \alpha} k_\alpha]\exp[ik(x-y)]
\label{61}
\eeq
With this, we can rewrite eq.(\ref{59}) in the form
\beq
<[j_0({\vec x},t),j_i({\vec y},t]> =
\lim_{\epsilon \to 0^+} I^\epsilon({\vec x} - {\vec y})
\label{62}
\eeq
with
\beq
I^\epsilon({\vec x}) = -2i \int \frac{d^{3}k}{({2\pi})^3}
k_{0}k_{i}
{\sin(k_0 \epsilon)} {\tilde F}(k)
\exp{(i \vec{k}.\vec{x})}
\label{63}
\eeq
where we have written $(k_\mu) = (k_o,k_i)$, $i=1, 2$. It will
be convenient to define
\beq
k'_0 = \epsilon k_0
\label{64}
\eeq
In terms of this new variable and using the explicit form for
${\tilde F}(k)$
given in \cite{BFO},\cite{4}, with $k = {(k_0^2 + \vec k^2)}^{1/2}$,
integral $I^\epsilon$ becomes
\beq
I^\epsilon({\vec x}) = - \frac{1}{8 \pi^2 \vert m \vert}
\frac{1}{\epsilon^2} \partial_i \int \frac{d^2k}{(2\pi)^2}
\exp{i (\vec{k}.\vec{x})} \int_0^\infty  dk'_0 k'_0 \sin k'_0 f(y)
\label{65}
\eeq
where
\beq
{f} (y) \;=\; \frac{1}{y} \,
\left[ 1 - \frac{(1-y)}{\sqrt{y}}
 \arcsin \frac{1}{\sqrt{1 + (1/y)}} \right]
\label{66}
\eeq
and we have defined
\beq
y = \frac{k^2}{4m^2} =
\frac{{k'}_0^2 + \epsilon^2 {\vec k}^2}{4 \epsilon^2 m^2}
\label{67}
\eeq
One can now see that  $ y \to \infty$ for
$\epsilon \to 0$ and fixed $m$. Then, expanding in powers of $1/y$
one has $f(y) \sim \pi/(2\sqrt y)$
and then using distribution theory to define the integral
over $k'_0$ one finds
\beq
<[j_0({\vec x},t),j_i({\vec y},t]>  =  -\frac{1}{8 \pi}
\lim_{\epsilon \to 0} \frac{1}{ \epsilon}
\partial_i\delta^{(2)}(\vec x - \vec y)
\label{68}
\eeq
This result for the
equal-time current commutator, evaluated
within the bosonized theory,
shows exactly the infinite Schwinger term that is found,
using the BJL method, for free
fermions in $d=3$ dimensions \cite{dVG}. As it happens in $d=4$
dimensions \cite{Ch}, we see
from eq.(\ref{68}) that the commutator at {\it unequal} times
is well defined: divergencies appear only when one takes the
equal-time limit.

One can evaluate also the next order vanishing in the equal-time
current commutator so as to compare it with the result from
the original fermion model reported in the
literature \cite{dVG}. The answer is \cite{4}
\begin{eqnarray}
<[j_0({\vec x},t),j_i({\vec y},t)]> & = & -\frac{1}{8 \pi}
\lim_{\epsilon \to 0}
\left(\frac{1}{ \epsilon}
 \partial_i\delta^{(2)}(\vec x - \vec y)  \right.\nonumber\\
 & & \left. -  \frac{\epsilon}{\Lambda} [4 m^2
 \partial_i\delta^{(2)}(\vec x - \vec y)
 - \frac{1}{2}
   \partial_i \Delta
 \delta^{(2)}(\vec x - \vec y)]\right)
\label{69}
\end{eqnarray}
where we have defined
\beq
\frac{1}{\Lambda} = \int_0^\infty dk'_0 \frac{1}{{k'}_0^2} \sin k'_0
\label{70}
\eeq
In order to compare with ref.\cite{dVG} where current
commutators were computed using dimensional regularization,
we define, coming back to the original variable
$k_0 = k'_0/\epsilon$
\beq
A[d] = \frac{1}{2}
\int d^{d-2}k_0 \frac{1}{k_0^2} sin k_0 \epsilon
\label{71}
\eeq
so that $A[d=3] = \epsilon/\Lambda$. One can now perform
the analytically continued integral to find, near $d=3$, the behavior
\beq
A[d] \sim - \epsilon \times \frac{\epsilon^{3-d}}{3-d}
\label{72}
\eeq
The same ambiguous result
for free fermions
is obtained in ref.\cite{dVG} near $d=3$. This ambiguity
can be however removed, the
pole in dimensional regularization corresponding as usual
to a logarithmic divergence. It is also interesting
to note that if one uses the nice approximation
${\tilde F}_{appr}$ for ${\tilde F}$ proposed
in ref.\cite{BFO}, one can well check the correctness
of our previous analysis \cite{4}.

{}From the analysis above, we see that not only the
infinite  Schwinger term, analogous to that
arising in  $d=4$ \cite{Ch}
is obtained in the bosonized
version of our $d=3$ fermion
theory but also the mass-dependent second term
as well as the triple derivative
third term, both vanishing in
the equal time limit.
Our analysis should be compared with that in ref.\cite{Ban}
where the fermionic commutator algebra is  inferred from
the Maxwell-Chern-Simons algebra for electric and magnetic
fields using a bosonization recipe  which is  valid
in the large mass limit.
One can see that in the large mass
regime,  terms depending on the product
$\epsilon m = \lambda $ will produce ambiguities according to the way
both limits ($\epsilon \to 0$ and $m \to \infty$) are taken
into account, a problem which is not present in the limit
of small masses.
To see this in more detail, let us come back to (\ref{63})
and consider the case in which $\lambda$ is kept fixed while
$\epsilon \to 0$ (so that $m \to \infty$). In this case,
taking the limit before integrating
out $k'_0$,
one finds for $I^\epsilon$
\beq
I^\epsilon({\vec x}) \sim \vert m \vert h(\lambda)
 \partial_i \delta^{(2)}({\vec x})
\label{73}
\eeq
where
\beq
h(\lambda) =
\frac{1}{2\pi} \int_0^\infty dz z\sin (2\lambda z) f(z)
\label{74}
\eeq
with $f$ given by eq.(\ref{66}).
Let us note that using the approximate ${\tilde F}_{appr}$
of ref.\cite{BFO} and taking the limit
after the exact integration over $k'_0$,
we recover the same behavior (\ref{73}).
We see that for  $\lambda = \epsilon m$ fixed, $h$ just gives
a normalization factor so that one reproduces from $I^\epsilon$
in the form (\ref{73}) a commutator algebra
at equal times and large mass that coincides with that
to be infered from
a Maxwell-Chern-Simons theory,
\beq
<[j_0({\vec x},t),j_i({\vec y},t]>
\longrightarrow
c \vert m \vert \partial_i \delta^{(2)}({\vec x} - {\vec y})
{}~~~(m \to \infty)
\label{75}
\eeq
with $c$ a normalization constant. Again, currents appearing
in the l.h.s. of eq.(\ref{75}) are bosonic currents
which can be written  in terms of the electric and magnetic fields
thus reproducing  the MCS gauge invariant algebra \cite{Jac},\cite{DJ}.
One should note however that the free fermion - MCS mapping
is valid in the large mass limit of the original fermionic
theory, this meaning the large-distances regime for fermion fields.
Since current commutators test the short-distance regime, one
should not take the MCS gauge-invariant algebra as a starting
point to reproduce the fermion current commutators.
\newpage
\section{The non-Abelian case in $d=3$}
In three dimensional space-time, the bosonic action (\ref{14})
takes the form
\begin{eqnarray}
\exp(-S_{bos}[A]) & = & \int \D b_\mu \D \bar c_\mu \D c_\mu
\exp \left( -tr \!\int \! \dv ( L[b] \pm \frac{i}{8\pi}
\varepsilon_{\mu \nu \alpha}
\bar c_\mu D_\nu[b] c_\alpha \right.
\nonumber \\
& & \left. \mp \frac{i}{16\pi}
(A_\mu - b_\mu)  {^*\!\!f}_\mu[b] ) \right)
\label{76}
\end{eqnarray}
Here ghost fields $\bar c_\mu$ and $c_\mu$ were introduced
to represent the Faddeev-Popov like determinant $\Delta[b]$.
 Again, we have written
\beq
tr \!\int \! \dv L[b] = - \log \det (\id + m + \bs)
\label{77}
\eeq
and we have chosen the arbitrary constant
 $\lambda$ appearing in (\ref{14}) so as to make contact with the
conventions of ref.\cite{7}, $\lambda = \frac{i}{16\pi}$. Moreover,
we have shifted the bosonic field $A_\mu \to A_\mu - b_\mu$
(this amounting to a trivial Jacobian) for reasons that will become
clear below.

 It was observed in ref.\cite{6}  that when $L[b]$ is
approximated by its first term in the $1/m$ expansion,
a set of BRST transformations can be defined so
that the corresponding BRST invariance allows to  obtain
the (approximate) bosonic action. We shall
explicitly prove here that this invariance is present
in (\ref{76}) where no
approximation for $L[b]$ is assumed.  To this end, we
introduce a set of auxiliary fields $h_\mu$
(taking values in the
Lie algebra of $G$),
$l$  and $\bar \chi$
so that one can rewrite (\ref{76}) in the
form
\beq
\exp(-S_{bos}[A])   =  \int \D b_\mu \D \bar c_\mu \D c_\mu
\D h_\mu \D l \D \bar \chi \exp (-S_{eff}[A,b,h,l,\bar c, c,
\bar \chi])
\label{cor}
\eeq
with
\begin{eqnarray}
S_{eff}[A,b,h,l,\bar c, c, \bar \chi] & = &
 tr \!\int \! \dv ( L[b-h]
\pm \frac{i}{8\pi}
\varepsilon_{\mu \nu \alpha}
\bar c_\mu D_\nu[b] c_\alpha
\nonumber\\
& &
 \mp \frac{i}{16\pi}
( (A_\mu - b_\mu)  {^*\!\!f}_\mu[b] + l h_\mu^2 - 2 \bar \chi
h_\mu c_\mu ) )  \label{11x}
\end{eqnarray}
where $\bar \chi$ is an anti-ghost field. Written in the form
(\ref{11x}), the bosonic action has a BRST invariance under the
following nilpotent off-shell BRST transformations %
\[
\delta \bar c_\mu = A_\mu - b_\mu , \;\;\;\;\;\;
\delta b_\mu = c_\mu , \;\;\;\;\;\; \delta A_\mu = c_\mu , \;\;\;\;\;\;
\delta c_\mu = 0 , \;\;\;\;\;\;
\delta \bar \chi = l
\]
\beq
\delta h_\mu = c_\mu , \;\;\;\;\;\; \delta l = 0
\label{17x}
\eeq

In view of this BRST invariance, one could add to $S_{eff}$  a BRST
exact form without changing the dynamics defined by $S_{bos}[A]$.
Exploiting this,
 we shall see that one can
factor out the $A_\mu$ dependence in
the r.h.s. of eq.(\ref{11x})  so that it completely decouples from
the path-integral over $b_\mu$ auxiliary and ghost fields
exactly as we did in the Abelian case. Although
complicated, this integral then becomes irrelevant for the definition
of the bosonic action for $A_\mu$. Indeed, let us add to $S_{eff}$
the BRST exact form $\delta G$,
\beq
S_{eff}[A,b,h,l,\bar c, c,\chi] \to S_{eff}[A,b,h,l,\bar c, c,\chi] +
\delta G[A, b, h, \bar c]
\label{20x}
\eeq
with
\beq
G[A, b, h, \bar c] = \mp \frac{i}{16 \pi} tr \!\int \! \dv
\varepsilon_{\mu \nu \alpha} \bar c_\mu H_{\nu \alpha}[A,b,h]
\label{21x}
\eeq
and $H_{\nu \alpha}[A,b,h]$ a functional to be determined in order to
produce the decoupling. Then, consider
the change of variables (analogous to (\ref{21}) for the Abelian case)
\beq
b_\mu = 2b'_\mu - A_\mu + V_\mu[A]
\label{18x}
\eeq
where $V_\mu[A]$ is some functional of $A_\mu$ changing covariantly
under gauge transformations,
\beq
V_\mu[A^g] = g^{-1} V_\mu[A] g
\label{19x}
\eeq
so that $b'_\mu$ is, like $A_\mu$ and $b_\mu$, a gauge field.
Integrating over $l$ in (\ref{cor}) and imposing the resulting
constraint, $h_\mu = 0$,  one
sees that if one imposes on $H_{\nu \alpha}[A,b,h]$ the condition
\begin{eqnarray}
\varepsilon _{\mu \nu \alpha}\!\!\int\!\!d^3y
 \!\left(
\frac{\delta H_{\nu \alpha}}{\delta  b^a_\rho (y)}\!+
\frac{\delta H_{\nu \alpha }}{\delta A^a_\rho (y)}\!+
\frac{\delta H_{\nu \alpha }}{\delta h^a_\rho (y)}
\right)\!c^a_\rho (y) \left|_{h=0}\right.\!=
\varepsilon _{\mu \nu \rho } [A_\nu\!-\! b_\nu\!-\!V_\nu[A], c_\rho]
 \nonumber\\
\label{28x}
\end{eqnarray}
then, when written in terms of the new $b'_\mu$ variable, the ghost
term becomes
\beq
S_{ghost}[b',c,\bar c] = \pm \frac{i}{8 \pi}
tr \! \int \!\dv \varepsilon_{\mu \nu \alpha}
\bar c_\mu D_\nu[b'] c_\alpha
\label{29x}
\eeq
so that its contribution is still $A_\mu$ independent. Then,
we can write the effective action in the form
\beq
S_{eff}[b',A] + S_{ghost}[b',c,\bar c]
\label{truc3}
\eeq
with
\begin{eqnarray}
S_{eff}[b',A] & = & \tilde{S}[b,A]
   \nonumber \\
& = & tr\!\int \! \dv \left( L[b]
\mp \frac{i}{16 \pi} (A_\mu - b_\mu)
(^*\!f_\mu[b] +  ^*\!H_\mu [A,b,0]) \right)
\label{30x}
\end{eqnarray}
where $^*\!H_\mu = \varepsilon_{\mu \nu \alpha} H_{\nu \alpha}$.

Condition (\ref{28x}) made the ghost term
independent of the bosonic field $A_\mu$. We shall now impose a
second constraint in order to completely  decouple the
auxiliary field $b'_\mu$ from $A_\mu$
in $S_{eff}$. Indeed, consider  the conditions
\beq
\frac{\delta ^2 S_{eff}[b',A]}{\delta A^a_\rho(y)
\delta \,b'^b_\sigma(x)} = 0
\label{31x}
\eeq
In terms of the original auxiliary field
 $b_\mu$  these equations read
\beq
\frac{\delta ^2 \tilde{S}[b,A]}{\delta A^a_\rho(y)\delta b^b
_\sigma(x)}  - \frac{\delta ^2 \tilde{S}[b,A]}{\delta b^a_\rho(y)
\delta b^b_\sigma(x)} + \int d^3u \frac{\delta
^2\tilde{S}[b,A]}{\delta b^c_\beta(u)\delta b^b_\sigma (x)} \,\frac
{\delta V^c_\beta(u)} {\delta A^a_\rho(y)} = 0
\label{32x}
\eeq
Eqs.(\ref{32x}) can be
easily written in terms of  $L$, $H$ and $V$ as a lengthy equation that
we shall omit here.

The strategy is now as follows: once a given approximate expression for the
fermion determinant is considered,
one should solve eq.(\ref{32x}) in order to
determine functionals $V$ in eq.(\ref{18x}) and $G$ in eq.(\ref{21x}),
 taking also in account the condition (\ref{28x}). In particular, if
one considers the $1/m$ expansion for the fermion determinant,
equations (\ref{28x}) and (\ref{32x}) should determine the form of $V$
and $G$ as a power expansion in $1/m$. In ref.\cite{Red} the $1/m$
expansion for the fermion determinant was shown to give
\beq
\ln \det (\id + m + \bs) =
  \pm \frac{i}{16\pi} S_{CS}[b] +
  I_{PC}[b] +
  O(\partial^2/m^2)  ,
\label{14x}
\eeq
where the Chern-Simons action $S_{CS}$ is given by
\beq
  S_{CS}[b] =
 \varepsilon_{\mu\nu\lambda} \tr \int\dv
 (
   f_{\mu \nu} b_{\lambda} -
   \frac{2}{3} b_{\mu}b_{\nu}b_{\lambda}
  )  .
\eeq
Concerning the parity conserving contributions, one has
\beq
I_{PC}[b] =
  - \frac{1}{24\pi  m} \tr\int\dv f^{\mu\nu} f_{\mu\nu}
  + \cdots  ,
\label{8f}
\eeq

To order zero in this expansion, solution of
eqs.(\ref{28x}),(\ref{32x}) is very simple. Indeed, in this case the
fermion determinant coincides with the CS action and one can easily
see that the solution is given by
\beq
V_\mu^{(0)}[A] = 0
\label{33x}
\eeq
\beq
G^{(0)}[A, b, h, \bar c]  =
\pm \frac{i}{16\pi} tr \!\int \! \dv \bar c_\mu \, (
 \frac{1}{2} \, ^*\!f_\mu[A] + \frac{1}{2} \, ^*\!f_\mu[b]
- 2 ^*\!D_{\mu\alpha}[A] h_\alpha
)
\label{34x}
\eeq
With this, the change of variables (\ref{18x}) takes the simple
form
\beq
b_\mu = 2 b'_\mu - A_\mu
\label{35x}
\eeq
and the decoupled effective action reads
\beq
S_{eff}^{(0)}[b,A,\bar c, c] = \mp \frac{i}{16\pi}
(2 S_{CS}[b'] - S_{CS}[A]) + S_{ghost}[b']
\label{36x}
\eeq
We then see that the path-integral  defining
the bosonic action $S_{bos}[A]$, factors out so that one
ends with a bosonic action in the form
\beq
S_{bos}^{(0)}[A] = \pm   \frac{i}{16\pi}
 S_{CS}[A]
\label{37x}
\eeq
as advanced in \cite{2},\cite{6}. Let us remark that
in finding the solution for $G$ one starts by writing the most
general form compatible with its dimensions,
\begin{eqnarray}
& & G^{(0)}[A, b, h, \bar c] = tr \!\int \! \dv
\varepsilon_{\mu \nu \alpha}
\bar c_\mu \, (
  d_1 b_\nu A_\alpha + d_2 A_\nu b_\alpha + d_3 b_\nu b_\alpha
+ d_4 A_\nu A_\alpha   \nonumber\\
& & + d_5  b_\nu h_\alpha  + d_6  h_\nu b_\alpha
+ d_7  A_\nu h_\alpha + d_8  h_\nu A_\alpha
+ d_9 \partial_\nu A_\alpha + d_{10} \partial_\nu b_\alpha
+ d_{11} \partial_\nu h_\alpha
) \nonumber \\
& & \label{48x}
\end{eqnarray}
All the arbitrary parameters $d_i$ are determined by imposing the
conditions (\ref{28x}) and (\ref{32x}) with $^*\!H_\mu$ transforming
covariantly (as $h_\mu$ does) under gauge transformations which
leads, together
with a gauge invariant action, to the solution (\ref{34x}).

To go further in the $1/m$ expansion one uses the next to the leading
order in the fermion determinant as given in eq.(\ref{14x}). Again,
starting from the general form of $G$ and after quite lengthy
calculations that we shall not reproduce here, one can find a unique
solution for $V_\mu$ and $H_{\nu \alpha}$ leading to a gauge
invariant action,
\beq
V_\mu^{(1)}[A] =  \pm \frac{2i}{3m} {^*\!f_\mu[A]}
\label{49x}
\eeq
\begin{eqnarray}
& & G^{(1)}[A,b,h,\bar c] = G^{(0)}[A,b,h,\bar c]
 \; \mp \; \frac{1}{96\pi m} tr \!\int \! \dv
\bar c_\mu \varepsilon_{\mu \nu \alpha}
\varepsilon_{\nu \rho \sigma}
\nonumber\\
& &
\left( \; \frac{1}{2} \; [ \; f_{\rho \sigma }[A-h] + 3 f_{\rho
\sigma }[b-h] - 2 D_\rho [A-h] (A_\sigma - b_\sigma) \; , \;
(A_\alpha - b_\alpha) \; ] \right.
\nonumber\\
& &
\left. + \; 4 \; [ \; f_{\rho \sigma }[A-h] \; , \; h_\alpha \; ] \;
\frac{}{} \right)
\label{50x}
\end{eqnarray}
The corresponding change of variables (\ref{18x}) takes now the
form
\beq
b_\mu = 2 b'_\mu - A_\mu \pm \frac{2i}{3m} {^*\!f_\mu}[A]
\label{399x}
\eeq
and the decoupled effective action reads
\beq
S_{eff}^{(1)}[b,A,\bar c, c] = S_{eff}^{(0)}[b,A,\bar c, c]
+ tr \!\int \! \dv \left(
\frac{1}{6\pi m} f_{\mu\nu}^2[b'] + \frac{1}{24\pi m} f_{\mu\nu}^2[A]
\right)
\label{400x}
\eeq
so that one can again  integrate out the completely
decoupled ghosts and $b'$ fields ending with the bosonic action
\beq
S_{bos}^{(1)}[A] =
\pm   \frac{i}{16\pi}
 S_{CS}[A] +  \frac{1}{24\pi m} tr \!\int \! \dv f_{\mu\nu}^2 [A]
\label{500x}
\eeq
This result extends to order $1/m$ the bosonization recipe presented
in refs. \cite{2},\cite{6}.

In this way, from the knowledge of the $1/m$ expansion of the fermion
determinant one can systematically find  order by order the
decoupling change of variables and construct the corresponding
action for the bosonic field $A_\mu$. One finds for the change of
variables
\beq
b_\mu = 2 b'_\mu - A_\mu  \pm \frac{2i}{3m} {^*\!f_\mu[A]}
+ \frac{1}{m^2} C^{(2)} D_\rho[A] f_{\mu \rho}[A] + \ldots
\label{510x}
\eeq
Here $C^{(2)}$ is a (dimensionless) constant to be determined from the
$1/m^2$ term in the fermion determinant expansion, which should be
proportional to $ {^*\!f_\mu} D_\rho f_{\rho \mu}$. Evidently,
finding the BRST exact form becomes more and more involved and so is
the form of the bosonic action which however, can be compactly
written as
\begin{eqnarray}
& & S_{bos}[A]  =  tr \!\int \! \dv  \left( \frac{}{}
L[-A + V[A]] \right.
\nonumber \\ & &
\left. \mp \frac{i}{16\pi}(2A_\mu - V_\mu[A]) ({^*\!f}_{\mu}[-A +
V[A]] + {^*\!H}_{\mu}[-A + V[A],A,0] ) \right)
\label{520x}
\end{eqnarray}
\vspace{0.5cm}

Let us end this section by writing the bosonization recipe for
the fermion current accompanying this result for the bosonic action.
{}From eq.(\ref{15}) we have, in $d=3$
\beq
\bar \psi^i \gamma_\mu t^a_{ij} \psi^j \to
\pm \frac{i}{8\pi}\varepsilon_{\mu \nu \alpha}
\partial_\nu A^a_\alpha
\label{109}
\eeq
\section{Wilson loops}
I will transcribe in this section some results obtained in \cite{2}
concerning the evaluation of Wilson loops in the framework of
our bosonization approach. In the Chern-Simons theory,
 they measure topological invariants determined by the linkings
of the loops and by the topology of the base manifold \cite{witten}.
For one loop $\Gamma$,
\beq
W[\Gamma] =
  \tr P \exp(i \oint_\Gamma dx^\mu\,A_\mu )
\label{wloop}
\eeq
where $P$ denotes the path ordering of the exponential, and the trace
is taken in the representation carried by the loop.  According to
the bosonization prescription, to relate this
operator to the fermionic theory we must express $W[\Gamma]$ in
terms of the field strength $F_{\mu\nu}$ rather than the potential
$A_\mu$.  In the abelian case this can always be done by means
of Stokes theorem.  As discussed in ref.~\cite{1}, this leads to
an explicit mapping between abelian Wilson loops and non-local
fermionic operators. Hence, in this way, the latter are related to
the linking of loops and thus probe the generalized statistics of
the external particles that propagate along those loops.  One way
to extend that analysis to the non-abelian case is to use the
non-abelian extension of Stokes theorem developed in~\cite{bralic}.
For an arbitrary loop $\Gamma = \partial\Sigma$, the boundary of a surface
$\Sigma$, one has
\beq
W[\partial\Sigma] = \tr P_t \exp\{ i\!\int_0^1 dt \int_0^1 ds
  \frac{\partial\Sigma^\mu}{\d s}\frac{\partial\Sigma^\nu}{\d t}
  W^{-1}[_s\Sigma(t)_0] F_{\mu\nu}(\Sigma(t,s)) W[_s\Sigma(t)_0] \}
\label{stokes}
\eeq
Here $\Sigma$ is looked upon as a sheet, that is, a one parameter
family of paths parametrized by $t$, $0 \leq t \leq 1$.  For each
$t$, $\Sigma(t)$ is a path, itself parametrized by $s$,
$0 \leq s \leq 1$, with fixed end-points:  $\partial\Sigma(t,s)/\d t = 0$
at $s = 0,1$.  For a given $t$, $_s\Sigma(t)_0$ denotes the segment
of the path $\Sigma(t)$ connecting the points $\Sigma(t,0)$ and
$\Sigma(t,s)$, and $W[_s\Sigma(t)_0]$ is the corresponding (open)
Wilson line.  Finally, $P_t$ in eq.~(\ref{stokes}) denotes ordering
of the $t$ integral, while the $s$ integral is not ordered (although
there is an $s$-ordering inside each $W[_s\Sigma(t)_0]$.)

In the abelian case the two open Wilson lines $W[_s\Sigma(t)_0]$
in eq.~(\ref{stokes}) cancel each other and one recovers the usual
Stokes theorem, involving only the gauge field strength.  In the
non-abelian case, however, the factors $W[_s\Sigma(t)_0]$ are needed
for gauge invariance, and introduce an explicit dependence of the
Wilson loop operator on the gauge potential $A_\mu$.  Thus, as
opposed to the abelian case, the non-abelian Wilson loop operator
cannot be mapped in a straightforward way to a fermionic operator
through the bosonization rule in eq.~(\ref{109}).

For planar loops this difficulty is only apparent.  Indeed, consider
$W[\partial\Sigma]$, with $\Sigma$ contained, say, in the $(1,2)$ plane.
Imposing the $A_3 = 0$ gauge condition, there is a remnant gauge
freedom for the $A_1$ and $A_2$ components in the $(1,2)$ plane,
which is the symmetry of a $2$-dimensional gauge theory in that
plane.  As discussed in~\cite{bralic}, one can use that gauge
symmetry, together with the freedom of parametrization of the
surface $\Sigma$, so the open Wilson line elements in the right hand
side of eq.~(\ref{stokes}) become the identity.  More precisely,
choosing the gauge condition $A_2 =0$ on the $\Sigma$-plane,
eq.~(\ref{stokes}) can be simplified to
\beq
W[\partial\Sigma] = \tr P_t \exp\{ i\!\int_0^1 dt \int_0^1 ds
  \frac{\partial\Sigma^\mu}{\d s}\frac{\partial\Sigma^\nu}{\d t}
  F_{\mu\nu}(\Sigma(t,s)) \}
\label{stokes-plane}
\eeq
provided that $\Sigma$ is parametrized so as to have $\partial\Sigma/\d t$
and $\partial\Sigma/\d s$ parallel to the $x_1$ and $x_2$ axis,
respectively.  This apparent breaking of rotational invariance,
which includes the presence of $t$-ordering but not of $s$-ordering,
is a consequence of the $A_2 =0$ gauge condition on the
$\Sigma$-plane, and will be removed by the functional integral over
the gauge fields.  One more ingredient is need: by an appropriate
addition of a BRST exact form, the bosonization rule (\ref{109}) can
be written in the covariant way
\beq
\bar \psi^i \gamma_\mu t^a_{ij} \psi^j \to
\pm \frac{i}{16\pi}\varepsilon_{\mu \nu \alpha}
F^a_{\nu\alpha}
\label{109x}
\eeq
Then, writing
\beq
\J[\Sigma] =
  \tr P_t \exp\{\pm 16\pi\!\int_0^1 dt \int_0^1 ds
  \frac{\partial\Sigma^\mu}{\d s}\frac{\partial\Sigma^\nu}{\d t}
  \epsilon_{\mu\nu\lambda} j_\lambda[\Sigma(t,s)] \}
\label{wsurface}
\eeq
with $j_\mu$ the fermionic current , the
bosonization formula  gives
\beq
\langle \J[\Sigma] \rangle_{ferm} =
  \langle W[\partial\Sigma] \rangle_{CS}
\label{bos-plane}
\eeq
where in the left hand side the subindex `ferm' stands for free
fermions.  This is the non-abelian generalization of the result
obtained in~\cite{1}.  It relates a suitably defined non-abelian
flux of the fermionic current through a flat surface, and the
Wilson loop associated to the boundary of that surface, with both
quantities in the same representation of the group.

It should be stressed that the bosonic side of this relation is,
by definition, independent of the surface $\Sigma$ and its
parametrization.  In the fermionic side, however, this is not
obvious.  The relation was derived assuming a flat surface $\Sigma$,
and it is tempting to assume that this may be extended to smooth
deformations away from the plane.  But more relevant is the apparent
breaking of rotational invariance in the fermionic side due to the
remaining $t$-ordering in eq.~(\ref{wsurface}). This should certainly
be expected to be taken care of by the particular parametrization
assumed above for $\Sigma$.  Indeed, one should expect that the very
need of a parametrization and of a matching ordering of the surface
integral of the fermionic current, is just a limitation of our
present analysis.  In addition, as is well known, the expectation
value of the Wilson loop is singular and must be regularized.  A
natural and consistent regularization scheme is provided by the
framing of the loop~\cite{witten}.  In the case of a
non-intersecting loop on a plane, considered here, that framing can
be chosen also as a plane loop, not intersecting itself nor the
original loop.  Again, it is not clear at this point how these
singularities in the bosonic side will show up in the (free)
fermionic side, and what role will the framing play from the
fermionic point of view.

It is natural to ask whether this analysis can be extended to
several loops and their possible linkings, as done in~\cite{1}
for the Abelian case.  In the bosonic side one is interested in
the expectation value
$\langle W[\Gamma_1] W[\Gamma_2] \rangle_{CS}$ or, better yet, in
the ratio
\beq
  \frac{\langle W[\Gamma_1] W[\Gamma_2] \rangle_{CS}}
       {\langle W[\Gamma_1] \rangle_{CS}
	    \langle W[\Gamma_2] \rangle_{CS}}
\label{two-loops}
\eeq

{
\begin{figure}
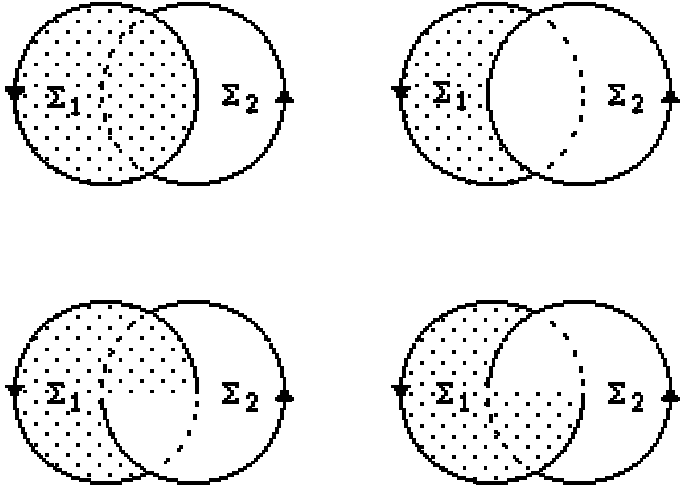

\baselineskip 0pt
\lineskip 0pt
\parskip 0pt
\parindent 0pt
\vbox{\vbox to 208pt{\vskip13pt
\hbox{\hskip43pt\vrule width12pt height1pt\hskip21pt\vrule width12pt
 height1pt\hskip101pt\vrule width12pt height1pt\hskip21pt\vrule width
12pt height1pt\hfil}
\hbox{\hskip39pt\vrule width4pt height1pt\hskip12pt\vrule width4pt
 height1pt\hskip13pt\vrule width4pt height1pt\hskip12pt\vrule width4pt
 height1pt\hskip93pt\vrule width4pt height1pt\hskip12pt\vrule width4pt
 height1pt\hskip13pt\vrule width4pt height1pt\hskip12pt\vrule width4pt
 height1pt\hfil}
\hbox{\hskip36pt\vrule width3pt height1pt\hskip20pt\vrule width3pt
 height1pt\hskip7pt\vrule width3pt height1pt\hskip20pt\vrule width3pt
 height1pt\hskip87pt\vrule width3pt height1pt\hskip20pt\vrule width3pt
 height1pt\hskip7pt\vrule width3pt height1pt\hskip20pt\vrule width3pt
 height1pt\hfil}
\hbox{\hskip34pt\vrule width2pt height1pt\hskip6pt\vrule width1pt
 height1pt\hskip7pt\vrule width1pt height1pt\hskip7pt\vrule width1pt
 height1pt\hskip3pt\vrule width2pt height1pt\hskip3pt\vrule width2pt
 height1pt\hskip26pt\vrule width2pt height1pt\hskip83pt\vrule width2pt
 height1pt\hskip4pt\vrule width1pt height1pt\hskip7pt\vrule width1pt
 height1pt\hskip7pt\vrule width1pt height1pt\hskip5pt\vrule width2pt
 height1pt\hskip3pt\vrule width2pt height1pt\hskip26pt\vrule width2pt
 height1pt\hfil}
\hbox{\hskip32pt\vrule width2pt height1pt\hskip30pt\vrule width3pt
 height1pt\hskip30pt\vrule width2pt height1pt\hskip79pt\vrule width2pt
 height1pt\hskip30pt\vrule width3pt height1pt\hskip30pt\vrule width2pt
 height1pt\hfil}
\hbox{\hskip30pt\vrule width2pt height1pt\hskip35pt\vrule width1pt
 height1pt\hskip31pt\vrule width2pt height1pt\hskip75pt\vrule width2pt
 height1pt\hskip31pt\vrule width1pt height1pt\hskip35pt\vrule width2pt
 height1pt\hfil}
\hbox{\hskip29pt\vrule width1pt height1pt\hskip38pt\vrule width1pt
 height1pt\hskip32pt\vrule width1pt height1pt\hskip73pt\vrule width1pt
 height1pt\hskip32pt\vrule width1pt height1pt\hskip38pt\vrule width1pt
 height1pt\hfil}
\hbox{\hskip27pt\vrule width2pt height1pt\hskip1pt\vrule width1pt
 height1pt\hskip7pt\vrule width1pt height1pt\hskip7pt\vrule width1pt
 height1pt\hskip7pt\vrule width1pt height1pt\hskip7pt\vrule width1pt
 height1pt\hskip6pt\vrule width2pt height1pt\hskip31pt\vrule width2pt
 height1pt\hskip69pt\vrule width2pt height1pt\hskip7pt\vrule width1pt
 height1pt\hskip7pt\vrule width1pt height1pt\hskip7pt\vrule width1pt
 height1pt\hskip7pt\vrule width2pt height1pt\hskip8pt\vrule width1pt
 height1pt\hskip31pt\vrule width2pt height1pt\hfil}
\hbox{\hskip26pt\vrule width1pt height1pt\hskip32pt\vrule width1pt
 height1pt\hskip11pt\vrule width1pt height1pt\hskip32pt\vrule width1pt
 height1pt\hskip67pt\vrule width1pt height1pt\hskip32pt\vrule width1pt
 height1pt\hskip11pt\vrule width1pt height1pt\hskip32pt\vrule width1pt
 height1pt\hfil}
\hbox{\hskip25pt\vrule width1pt height1pt\hskip32pt\vrule width1pt
 height1pt\hskip13pt\vrule width1pt height1pt\hskip32pt\vrule width1pt
 height1pt\hskip65pt\vrule width1pt height1pt\hskip32pt\vrule width1pt
 height1pt\hskip13pt\vrule width1pt height1pt\hskip32pt\vrule width1pt
 height1pt\hfil}
\hbox{\hskip24pt\vrule width1pt height1pt\hskip48pt\vrule width1pt
 height1pt\hskip32pt\vrule width1pt height1pt\hskip96pt\vrule width1pt
 height1pt\hskip48pt\vrule width1pt height1pt\hfil}
\hbox{\hskip23pt\vrule width1pt height1pt\hskip2pt\vrule width1pt
 height1pt\hskip7pt\vrule width1pt height1pt\hskip7pt\vrule width1pt
 height1pt\hskip7pt\vrule width1pt height1pt\hskip7pt\vrule width1pt
 height1pt\hskip7pt\vrule width1pt height1pt\hskip7pt\vrule width1pt
 height1pt\hskip32pt\vrule width1pt height1pt\hskip61pt\vrule width1pt
 height1pt\hskip8pt\vrule width1pt height1pt\hskip7pt\vrule width1pt
 height1pt\hskip7pt\vrule width1pt height1pt\hskip7pt\vrule width1pt
 height1pt\hskip50pt\vrule width1pt height1pt\hfil}
\hbox{\hskip22pt\vrule width1pt height1pt\hskip32pt\vrule width1pt
 height1pt\hskip19pt\vrule width1pt height1pt\hskip32pt\vrule width1pt
 height1pt\hskip59pt\vrule width1pt height1pt\hskip32pt\vrule width1pt
 height1pt\hskip52pt\vrule width1pt height1pt\hfil}
\hbox{\hskip54pt\vrule width1pt height1pt\hfil}
\hbox{\hskip222pt\vrule width1pt height1pt\hfil}
\hbox{\hskip20pt\vrule width1pt height1pt\hskip1pt\vrule width1pt
 height1pt\hskip7pt\vrule width1pt height1pt\hskip7pt\vrule width1pt
 height1pt\hskip7pt\vrule width1pt height1pt\hskip7pt\vrule width1pt
 height1pt\hskip7pt\vrule width1pt height1pt\hskip7pt\vrule width1pt
 height1pt\hskip6pt\vrule width1pt height1pt\hskip32pt\vrule width1pt
 height1pt\hskip55pt\vrule width1pt height1pt\hskip7pt\vrule width1pt
 height1pt\hskip7pt\vrule width1pt height1pt\hskip7pt\vrule width1pt
 height1pt\hskip8pt\vrule width1pt height1pt\hskip23pt\vrule width1pt
 height1pt\hskip32pt\vrule width1pt height1pt\hfil}
\vskip2pt
\hbox{\hskip51pt\vrule width1pt height1pt\hfil}
\hbox{\hskip26pt\vrule width1pt height1pt\hskip7pt\vrule width1pt
 height1pt\hskip7pt\vrule width1pt height1pt\hskip15pt\vrule width1pt
 height1pt\hskip7pt\vrule width1pt height1pt\hskip7pt\vrule width1pt
 height1pt\hskip95pt\vrule width1pt height1pt\hskip7pt\vrule width1pt
 height1pt\hskip7pt\vrule width1pt height1pt\hskip7pt\vrule width1pt
 height1pt\hfil}
\vskip3pt
\hbox{\hskip22pt\vrule width1pt height1pt\hskip7pt\vrule width1pt
 height1pt\hskip7pt\vrule width1pt height1pt\hskip7pt\vrule width1pt
 height1pt\hskip7pt\vrule width1pt height1pt\hskip7pt\vrule width1pt
 height1pt\hskip7pt\vrule width1pt height1pt\hskip7pt\vrule width1pt
 height1pt\hskip87pt\vrule width1pt height1pt\hskip7pt\vrule width1pt
 height1pt\hskip7pt\vrule width1pt height1pt\hskip7pt\vrule width1pt
 height1pt\hfil}
\hbox{\hskip49pt\vrule width1pt height1pt\hfil}
\vskip2pt
\hbox{\hskip18pt\vrule width1pt height1pt\hskip7pt\vrule width1pt
 height1pt\hskip7pt\vrule width1pt height1pt\hskip7pt\vrule width1pt
 height1pt\hskip7pt\vrule width1pt height1pt\hskip7pt\vrule width1pt
 height1pt\hskip7pt\vrule width1pt height1pt\hskip7pt\vrule width1pt
 height1pt\hskip87pt\vrule width1pt height1pt\hskip7pt\vrule width1pt
 height1pt\hskip7pt\vrule width1pt height1pt\hskip7pt\vrule width1pt
 height1pt\hfil}
\vskip2pt
\hbox{\hskip173pt\vrule width6pt height1pt\hfil}
\hbox{\hskip22pt\vrule width1pt height1pt\hskip15pt\vrule width1pt
 height1pt\hskip7pt\vrule width1pt height1pt\hskip7pt\vrule width1pt
 height1pt\hskip7pt\vrule width1pt height1pt\hskip7pt\vrule width1pt
 height1pt\hskip7pt\vrule width1pt height1pt\hskip14pt\vrule width6pt
 height1pt\hskip67pt\vrule width1pt height1pt\hskip7pt\vrule width1pt
 height1pt\hskip3pt\vrule width1pt height1pt\hskip3pt\vrule width1pt
 height1pt\hskip7pt\vrule width1pt height1pt\hskip48pt\vrule width6pt
 height1pt\hfil}
\hbox{\hskip11pt\vrule width3pt height1pt\hskip13pt\vrule width6pt
 height1pt\hskip61pt\vrule width1pt height1pt\hskip3pt\vrule width1pt
 height1pt\hskip58pt\vrule width3pt height1pt\hskip2pt\vrule width2pt
 height1pt\hskip11pt\vrule width1pt height1pt\hskip64pt\vrule width1pt
 height1pt\hskip3pt\vrule width1pt height1pt\hfil}
\hbox{\hskip11pt\vrule width3pt height1pt\hskip14pt\vrule width1pt
 height1pt\hskip3pt\vrule width1pt height1pt\hskip62pt\vrule width1pt
 height1pt\hskip62pt\vrule width2pt height1pt\hskip2pt\vrule width1pt
 height1pt\hskip78pt\vrule width1pt height1pt\hfil}
\hbox{\hskip12pt\vrule width2pt height1pt\hskip15pt\vrule width1pt
 height1pt\hfil}
\hbox{\hskip18pt\vrule width1pt height1pt\hskip7pt\vrule width1pt
 height1pt\hskip15pt\vrule width1pt height1pt\hskip7pt\vrule width1pt
 height1pt\hskip7pt\vrule width1pt height1pt\hskip7pt\vrule width1pt
 height1pt\hskip7pt\vrule width1pt height1pt\hskip7pt\vrule width1pt
 height1pt\hskip20pt\vrule width2pt height1pt\hskip57pt\vrule width1pt
 height1pt\hskip7pt\vrule width1pt height1pt\hskip4pt\vrule width1pt
 height1pt\hskip6pt\vrule width1pt height1pt\hskip3pt\vrule width1pt
 height1pt\hskip62pt\vrule width2pt height1pt\hfil}
\hbox{\hskip95pt\vrule width1pt height1pt\hskip6pt\vrule width1pt
 height1pt\hskip138pt\vrule width1pt height1pt\hskip6pt\vrule width1pt
 height1pt\hfil}
\hbox{\hskip29pt\vrule width1pt height1pt\hskip6pt\vrule width1pt
 height1pt\hskip76pt\vrule width1pt height1pt\hskip5pt\vrule width1pt
 height1pt\hskip53pt\vrule width1pt height1pt\hskip1pt\vrule width3pt
 height1pt\hskip81pt\vrule width1pt height1pt\hskip5pt\vrule width1pt
 height1pt\hfil}
\hbox{\hskip93pt\vrule width1pt height1pt\hskip1pt\vrule width3pt
 height1pt\hskip141pt\vrule width1pt height1pt\hskip1pt\vrule width3pt
 height1pt\hfil}
\hbox{\hskip22pt\vrule width1pt height1pt\hskip4pt\vrule width1pt
 height1pt\hskip1pt\vrule width3pt height1pt\hskip14pt\vrule width1pt
 height1pt\hskip7pt\vrule width1pt height1pt\hskip7pt\vrule width1pt
 height1pt\hskip7pt\vrule width1pt height1pt\hskip7pt\vrule width1pt
 height1pt\hskip87pt\vrule width1pt height1pt\hskip7pt\vrule width1pt
 height1pt\hskip15pt\vrule width1pt height1pt\hfil}
\hbox{\hskip182pt\vrule width1pt height1pt\hskip1pt\vrule width1pt
 height1pt\hfil}
\hbox{\hskip102pt\vrule width1pt height1pt\hskip1pt\vrule width2pt
 height1pt\hskip142pt\vrule width1pt height1pt\hskip1pt\vrule width2pt
 height1pt\hfil}
\hbox{\hskip36pt\vrule width1pt height1pt\hskip1pt\vrule width1pt
 height1pt\hfil}
\hbox{\hskip18pt\vrule width1pt height1pt\hskip7pt\vrule width1pt
 height1pt\hskip7pt\vrule width1pt height1pt\hskip7pt\vrule width1pt
 height1pt\hskip7pt\vrule width1pt height1pt\hskip7pt\vrule width1pt
 height1pt\hskip7pt\vrule width1pt height1pt\hskip7pt\vrule width1pt
 height1pt\hskip87pt\vrule width1pt height1pt\hskip7pt\vrule width1pt
 height1pt\hskip7pt\vrule width1pt height1pt\hskip7pt\vrule width1pt
 height1pt\hfil}
\vskip3pt
\hbox{\hskip22pt\vrule width1pt height1pt\hskip7pt\vrule width1pt
 height1pt\hskip7pt\vrule width1pt height1pt\hskip7pt\vrule width1pt
 height1pt\hskip7pt\vrule width1pt height1pt\hskip7pt\vrule width1pt
 height1pt\hskip7pt\vrule width1pt height1pt\hskip7pt\vrule width1pt
 height1pt\hskip87pt\vrule width1pt height1pt\hskip7pt\vrule width1pt
 height1pt\hskip7pt\vrule width1pt height1pt\hskip7pt\vrule width1pt
 height1pt\hfil}
\vskip1pt
\hbox{\hskip226pt\vrule width1pt height1pt\hfil}
\hbox{\hskip225pt\vrule width1pt height1pt\hfil}
\hbox{\hskip26pt\vrule width1pt height1pt\hskip7pt\vrule width1pt
 height1pt\hskip7pt\vrule width1pt height1pt\hskip7pt\vrule width1pt
 height1pt\hskip7pt\vrule width1pt height1pt\hskip7pt\vrule width1pt
 height1pt\hskip7pt\vrule width1pt height1pt\hskip95pt\vrule width1pt
 height1pt\hskip7pt\vrule width1pt height1pt\hskip7pt\vrule width1pt
 height1pt\hskip7pt\vrule width1pt height1pt\hfil}
\vskip1pt
\hbox{\hskip52pt\vrule width1pt height1pt\hfil}
\hbox{\hskip20pt\vrule width1pt height1pt\hskip32pt\vrule width1pt
 height1pt\hskip23pt\vrule width1pt height1pt\hskip32pt\vrule width1pt
 height1pt\hskip55pt\vrule width1pt height1pt\hskip32pt\vrule width1pt
 height1pt\hskip56pt\vrule width1pt height1pt\hfil}
\hbox{\hskip21pt\vrule width2pt height1pt\hskip7pt\vrule width1pt
 height1pt\hskip7pt\vrule width1pt height1pt\hskip7pt\vrule width1pt
 height1pt\hskip15pt\vrule width1pt height1pt\hskip7pt\vrule width1pt
 height1pt\hskip103pt\vrule width1pt height1pt\hskip7pt\vrule width1pt
 height1pt\hskip7pt\vrule width1pt height1pt\hskip7pt\vrule width1pt
 height1pt\hfil}
\hbox{\hskip21pt\vrule width1pt height1pt\hskip32pt\vrule width1pt
 height1pt\hskip167pt\vrule width1pt height1pt\hfil}
\hbox{\hskip22pt\vrule width1pt height1pt\hskip52pt\vrule width1pt
 height1pt\hskip32pt\vrule width1pt height1pt\hskip59pt\vrule width1pt
 height1pt\hskip32pt\vrule width1pt height1pt\hskip19pt\vrule width1pt
 height1pt\hskip32pt\vrule width1pt height1pt\hfil}
\hbox{\hskip23pt\vrule width1pt height1pt\hskip50pt\vrule width1pt
 height1pt\hskip32pt\vrule width1pt height1pt\hskip61pt\vrule width1pt
 height1pt\hskip83pt\vrule width1pt height1pt\hfil}
\hbox{\hskip24pt\vrule width1pt height1pt\hskip1pt\vrule width1pt
 height1pt\hskip7pt\vrule width1pt height1pt\hskip7pt\vrule width1pt
 height1pt\hskip7pt\vrule width1pt height1pt\hskip7pt\vrule width1pt
 height1pt\hskip7pt\vrule width1pt height1pt\hskip6pt\vrule width1pt
 height1pt\hskip32pt\vrule width1pt height1pt\hskip63pt\vrule width1pt
 height1pt\hskip7pt\vrule width1pt height1pt\hskip7pt\vrule width1pt
 height1pt\hskip7pt\vrule width1pt height1pt\hskip8pt\vrule width1pt
 height1pt\hskip48pt\vrule width1pt height1pt\hfil}
\hbox{\hskip25pt\vrule width1pt height1pt\hskip46pt\vrule width1pt
 height1pt\hskip32pt\vrule width1pt height1pt\hskip65pt\vrule width1pt
 height1pt\hskip32pt\vrule width1pt height1pt\hskip46pt\vrule width1pt
 height1pt\hfil}
\hbox{\hskip26pt\vrule width1pt height1pt\hskip32pt\vrule width1pt
 height1pt\hskip11pt\vrule width1pt height1pt\hskip32pt\vrule width1pt
 height1pt\hskip67pt\vrule width1pt height1pt\hskip32pt\vrule width1pt
 height1pt\hskip11pt\vrule width1pt height1pt\hskip32pt\vrule width1pt
 height1pt\hfil}
\hbox{\hskip27pt\vrule width2pt height1pt\hskip31pt\vrule width1pt
 height1pt\hskip8pt\vrule width2pt height1pt\hskip31pt\vrule width2pt
 height1pt\hskip69pt\vrule width2pt height1pt\hskip31pt\vrule width2pt
 height1pt\hskip8pt\vrule width1pt height1pt\hskip31pt\vrule width2pt
 height1pt\hfil}
\hbox{\hskip29pt\vrule width2pt height1pt\hskip7pt\vrule width1pt
 height1pt\hskip7pt\vrule width1pt height1pt\hskip7pt\vrule width1pt
 height1pt\hskip7pt\vrule width1pt height1pt\hskip5pt\vrule width1pt
 height1pt\hskip32pt\vrule width1pt height1pt\hskip73pt\vrule width1pt
 height1pt\hskip6pt\vrule width1pt height1pt\hskip7pt\vrule width1pt
 height1pt\hskip7pt\vrule width1pt height1pt\hskip7pt\vrule width1pt
 height1pt\hskip1pt\vrule width1pt height1pt\hskip38pt\vrule width1pt
 height1pt\hfil}
\hbox{\hskip30pt\vrule width2pt height1pt\hskip35pt\vrule width1pt
 height1pt\hskip31pt\vrule width2pt height1pt\hskip75pt\vrule width2pt
 height1pt\hskip31pt\vrule width1pt height1pt\hskip35pt\vrule width2pt
 height1pt\hfil}
\hbox{\hskip32pt\vrule width2pt height1pt\hskip30pt\vrule width3pt
 height1pt\hskip30pt\vrule width2pt height1pt\hskip79pt\vrule width2pt
 height1pt\hskip30pt\vrule width3pt height1pt\hskip30pt\vrule width2pt
 height1pt\hfil}
\hbox{\hskip34pt\vrule width2pt height1pt\hskip26pt\vrule width2pt
 height1pt\hskip3pt\vrule width2pt height1pt\hskip26pt\vrule width2pt
 height1pt\hskip83pt\vrule width2pt height1pt\hskip26pt\vrule width2pt
 height1pt\hskip3pt\vrule width2pt height1pt\hskip26pt\vrule width2pt
 height1pt\hfil}
\hbox{\hskip36pt\vrule width3pt height1pt\hskip11pt\vrule width1pt
 height1pt\hskip7pt\vrule width4pt height1pt\hskip7pt\vrule width3pt
 height1pt\hskip20pt\vrule width3pt height1pt\hskip87pt\vrule width3pt
 height1pt\hskip9pt\vrule width1pt height1pt\hskip10pt\vrule width3pt
 height1pt\hskip7pt\vrule width3pt height1pt\hskip20pt\vrule width3pt
 height1pt\hfil}
\hbox{\hskip39pt\vrule width3pt height1pt\hskip13pt\vrule width4pt
 height1pt\hskip13pt\vrule width4pt height1pt\hskip12pt\vrule width4pt
 height1pt\hskip93pt\vrule width1pt height1pt\hskip1pt\vrule width2pt
 height1pt\hskip12pt\vrule width1pt height1pt\hskip1pt\vrule width2pt
 height1pt\hskip13pt\vrule width4pt height1pt\hskip12pt\vrule width4pt
 height1pt\hfil}
\hbox{\hskip43pt\vrule width12pt height1pt\hskip21pt\vrule width12pt
 height1pt\hskip101pt\vrule width12pt height1pt\hskip21pt\vrule width
12pt height1pt\hfil}
\vskip43pt
\hbox{\hskip43pt\vrule width12pt height1pt\hskip21pt\vrule width12pt
 height1pt\hskip101pt\vrule width12pt height1pt\hskip21pt\vrule width
12pt height1pt\hfil}
\hbox{\hskip39pt\vrule width4pt height1pt\hskip12pt\vrule width4pt
 height1pt\hskip13pt\vrule width4pt height1pt\hskip12pt\vrule width4pt
 height1pt\hskip93pt\vrule width4pt height1pt\hskip12pt\vrule width4pt
 height1pt\hskip13pt\vrule width4pt height1pt\hskip12pt\vrule width4pt
 height1pt\hfil}
\hbox{\hskip36pt\vrule width3pt height1pt\hskip3pt\vrule width1pt
 height1pt\hskip7pt\vrule width1pt height1pt\hskip7pt\vrule width4pt
 height1pt\hskip7pt\vrule width3pt height1pt\hskip20pt\vrule width3pt
 height1pt\hskip87pt\vrule width3pt height1pt\hskip20pt\vrule width3pt
 height1pt\hskip7pt\vrule width3pt height1pt\hskip20pt\vrule width3pt
 height1pt\hfil}
\hbox{\hskip34pt\vrule width2pt height1pt\hskip26pt\vrule width2pt
 height1pt\hskip3pt\vrule width2pt height1pt\hskip26pt\vrule width2pt
 height1pt\hskip83pt\vrule width2pt height1pt\hskip6pt\vrule width1pt
 height1pt\hskip7pt\vrule width1pt height1pt\hskip7pt\vrule width1pt
 height1pt\hskip3pt\vrule width2pt height1pt\hskip3pt\vrule width2pt
 height1pt\hskip26pt\vrule width2pt height1pt\hfil}
\hbox{\hskip32pt\vrule width2pt height1pt\hskip30pt\vrule width3pt
 height1pt\hskip30pt\vrule width2pt height1pt\hskip79pt\vrule width2pt
 height1pt\hskip30pt\vrule width3pt height1pt\hskip30pt\vrule width2pt
 height1pt\hfil}
\hbox{\hskip30pt\vrule width2pt height1pt\hskip35pt\vrule width1pt
 height1pt\hskip31pt\vrule width2pt height1pt\hskip75pt\vrule width2pt
 height1pt\hskip31pt\vrule width1pt height1pt\hskip35pt\vrule width2pt
 height1pt\hfil}
\hbox{\hskip29pt\vrule width2pt height1pt\hskip7pt\vrule width1pt
 height1pt\hskip7pt\vrule width1pt height1pt\hskip7pt\vrule width1pt
 height1pt\hskip7pt\vrule width1pt height1pt\hskip5pt\vrule width1pt
 height1pt\hskip32pt\vrule width1pt height1pt\hskip73pt\vrule width1pt
 height1pt\hskip32pt\vrule width1pt height1pt\hskip38pt\vrule width1pt
 height1pt\hfil}
\hbox{\hskip27pt\vrule width2pt height1pt\hskip40pt\vrule width2pt
 height1pt\hskip31pt\vrule width2pt height1pt\hskip69pt\vrule width2pt
 height1pt\hskip1pt\vrule width1pt height1pt\hskip7pt\vrule width1pt
 height1pt\hskip7pt\vrule width1pt height1pt\hskip7pt\vrule width1pt
 height1pt\hskip5pt\vrule width2pt height1pt\hskip40pt\vrule width2pt
 height1pt\hfil}
\hbox{\hskip26pt\vrule width1pt height1pt\hskip32pt\vrule width1pt
 height1pt\hskip11pt\vrule width1pt height1pt\hskip32pt\vrule width1pt
 height1pt\hskip67pt\vrule width1pt height1pt\hskip32pt\vrule width1pt
 height1pt\hskip11pt\vrule width1pt height1pt\hskip32pt\vrule width1pt
 height1pt\hfil}
\hbox{\hskip25pt\vrule width1pt height1pt\hskip32pt\vrule width1pt
 height1pt\hskip13pt\vrule width1pt height1pt\hskip32pt\vrule width1pt
 height1pt\hskip65pt\vrule width1pt height1pt\hskip32pt\vrule width1pt
 height1pt\hskip13pt\vrule width1pt height1pt\hskip32pt\vrule width1pt
 height1pt\hfil}
\hbox{\hskip24pt\vrule width1pt height1pt\hskip1pt\vrule width1pt
 height1pt\hskip7pt\vrule width1pt height1pt\hskip7pt\vrule width1pt
 height1pt\hskip7pt\vrule width1pt height1pt\hskip6pt\vrule width1pt
 height1pt\hskip8pt\vrule width1pt height1pt\hskip6pt\vrule width1pt
 height1pt\hskip32pt\vrule width1pt height1pt\hskip63pt\vrule width1pt
 height1pt\hskip32pt\vrule width1pt height1pt\hskip48pt\vrule width1pt
 height1pt\hfil}
\hbox{\hskip23pt\vrule width1pt height1pt\hskip50pt\vrule width1pt
 height1pt\hskip32pt\vrule width1pt height1pt\hskip61pt\vrule width1pt
 height1pt\hskip2pt\vrule width1pt height1pt\hskip7pt\vrule width1pt
 height1pt\hskip7pt\vrule width1pt height1pt\hskip7pt\vrule width1pt
 height1pt\hskip5pt\vrule width1pt height1pt\hskip50pt\vrule width1pt
 height1pt\hfil}
\hbox{\hskip22pt\vrule width1pt height1pt\hskip52pt\vrule width1pt
 height1pt\hskip32pt\vrule width1pt height1pt\hskip59pt\vrule width1pt
 height1pt\hskip32pt\vrule width1pt height1pt\hskip52pt\vrule width1pt
 height1pt\hfil}
\hbox{\hskip222pt\vrule width1pt height1pt\hfil}
\hbox{\hskip22pt\vrule width1pt height1pt\hskip7pt\vrule width1pt
 height1pt\hskip7pt\vrule width1pt height1pt\hskip7pt\vrule width1pt
 height1pt\hskip7pt\vrule width1pt height1pt\hskip7pt\vrule width1pt
 height1pt\hskip7pt\vrule width1pt height1pt\hskip152pt\vrule width1pt
 height1pt\hfil}
\hbox{\hskip20pt\vrule width1pt height1pt\hskip32pt\vrule width1pt
 height1pt\hskip23pt\vrule width1pt height1pt\hskip32pt\vrule width1pt
 height1pt\hskip55pt\vrule width1pt height1pt\hskip1pt\vrule width1pt
 height1pt\hskip7pt\vrule width1pt height1pt\hskip7pt\vrule width1pt
 height1pt\hskip7pt\vrule width1pt height1pt\hskip6pt\vrule width1pt
 height1pt\hskip56pt\vrule width1pt height1pt\hfil}
\vskip2pt
\hbox{\hskip26pt\vrule width1pt height1pt\hskip7pt\vrule width1pt
 height1pt\hskip7pt\vrule width1pt height1pt\hskip7pt\vrule width1pt
 height1pt\hskip7pt\vrule width1pt height1pt\hskip7pt\vrule width1pt
 height1pt\hskip7pt\vrule width1pt height1pt\hfil}
\hbox{\hskip51pt\vrule width1pt height1pt\hskip120pt\vrule width1pt
 height1pt\hskip7pt\vrule width1pt height1pt\hskip7pt\vrule width1pt
 height1pt\hfil}
\hbox{\hskip50pt\vrule width1pt height1pt\hfil}
\vskip1pt
\hbox{\hskip22pt\vrule width1pt height1pt\hskip7pt\vrule width1pt
 height1pt\hskip7pt\vrule width1pt height1pt\hskip7pt\vrule width1pt
 height1pt\hskip7pt\vrule width1pt height1pt\hskip7pt\vrule width1pt
 height1pt\hskip7pt\vrule width1pt height1pt\hskip7pt\vrule width1pt
 height1pt\hfil}
\hbox{\hskip168pt\vrule width1pt height1pt\hskip7pt\vrule width1pt
 height1pt\hskip7pt\vrule width1pt height1pt\hskip7pt\vrule width1pt
 height1pt\hfil}
\vskip2pt
\hbox{\hskip18pt\vrule width1pt height1pt\hskip7pt\vrule width1pt
 height1pt\hskip7pt\vrule width1pt height1pt\hskip7pt\vrule width1pt
 height1pt\hskip7pt\vrule width1pt height1pt\hskip7pt\vrule width1pt
 height1pt\hskip7pt\vrule width1pt height1pt\hskip7pt\vrule width1pt
 height1pt\hfil}
\hbox{\hskip164pt\vrule width1pt height1pt\hskip7pt\vrule width1pt
 height1pt\hskip7pt\vrule width1pt height1pt\hskip7pt\vrule width1pt
 height1pt\hfil}
\vskip2pt
\hbox{\hskip22pt\vrule width1pt height1pt\hskip15pt\vrule width1pt
 height1pt\hskip7pt\vrule width2pt height1pt\hskip6pt\vrule width1pt
 height1pt\hskip7pt\vrule width1pt height1pt\hskip7pt\vrule width1pt
 height1pt\hskip7pt\vrule width1pt height1pt\hfil}
\hbox{\hskip27pt\vrule width6pt height1pt\hskip60pt\vrule width6pt
 height1pt\hskip69pt\vrule width1pt height1pt\hskip5pt\vrule width6pt
 height1pt\hskip4pt\vrule width1pt height1pt\hskip7pt\vrule width1pt
 height1pt\hskip46pt\vrule width6pt height1pt\hfil}
\hbox{\hskip11pt\vrule width3pt height1pt\hskip2pt\vrule width2pt
 height1pt\hskip10pt\vrule width1pt height1pt\hskip3pt\vrule width1pt
 height1pt\hskip61pt\vrule width1pt height1pt\hskip3pt\vrule width1pt
 height1pt\hskip58pt\vrule width3pt height1pt\hskip2pt\vrule width2pt
 height1pt\hskip11pt\vrule width1pt height1pt\hskip3pt\vrule width1pt
 height1pt\hskip60pt\vrule width1pt height1pt\hskip3pt\vrule width1pt
 height1pt\hfil}
\hbox{\hskip12pt\vrule width2pt height1pt\hskip2pt\vrule width1pt
 height1pt\hskip12pt\vrule width1pt height1pt\hskip65pt\vrule width1pt
 height1pt\hskip62pt\vrule width2pt height1pt\hskip2pt\vrule width1pt
 height1pt\hskip13pt\vrule width1pt height1pt\hskip64pt\vrule width1pt
 height1pt\hfil}
\hbox{\hskip18pt\vrule width1pt height1pt\hskip7pt\vrule width1pt
 height1pt\hskip15pt\vrule width1pt height1pt\hskip7pt\vrule width1pt
 height1pt\hskip7pt\vrule width1pt height1pt\hskip7pt\vrule width1pt
 height1pt\hskip7pt\vrule width1pt height1pt\hfil}
\hbox{\hskip103pt\vrule width2pt height1pt\hskip59pt\vrule width1pt
 height1pt\hskip7pt\vrule width1pt height1pt\hskip15pt\vrule width1pt
 height1pt\hskip7pt\vrule width1pt height1pt\hskip7pt\vrule width1pt
 height1pt\hskip7pt\vrule width1pt height1pt\hskip7pt\vrule width1pt
 height1pt\hskip28pt\vrule width2pt height1pt\hfil}
\hbox{\hskip29pt\vrule width1pt height1pt\hskip6pt\vrule width1pt
 height1pt\hskip58pt\vrule width1pt height1pt\hskip6pt\vrule width1pt
 height1pt\hskip73pt\vrule width1pt height1pt\hskip6pt\vrule width1pt
 height1pt\hskip57pt\vrule width1pt height1pt\hskip6pt\vrule width1pt
 height1pt\hfil}
\hbox{\hskip113pt\vrule width1pt height1pt\hskip5pt\vrule width1pt
 height1pt\hskip139pt\vrule width1pt height1pt\hskip5pt\vrule width1pt
 height1pt\hfil}
\hbox{\hskip22pt\vrule width1pt height1pt\hskip4pt\vrule width1pt
 height1pt\hskip1pt\vrule width3pt height1pt\hskip14pt\vrule width1pt
 height1pt\hskip46pt\vrule width1pt height1pt\hskip1pt\vrule width3pt
 height1pt\hskip76pt\vrule width1pt height1pt\hskip1pt\vrule width3pt
 height1pt\hskip60pt\vrule width1pt height1pt\hskip1pt\vrule width3pt
 height1pt\hfil}
\hbox{\hskip168pt\vrule width1pt height1pt\hskip23pt\vrule width1pt
 height1pt\hskip7pt\vrule width1pt height1pt\hskip7pt\vrule width1pt
 height1pt\hskip7pt\vrule width1pt height1pt\hskip7pt\vrule width1pt
 height1pt\hfil}
\vskip1pt
\hbox{\hskip36pt\vrule width1pt height1pt\hskip1pt\vrule width1pt
 height1pt\hskip63pt\vrule width1pt height1pt\hskip1pt\vrule width2pt
 height1pt\hskip77pt\vrule width1pt height1pt\hskip1pt\vrule width1pt
 height1pt\hskip62pt\vrule width1pt height1pt\hskip1pt\vrule width2pt
 height1pt\hfil}
\hbox{\hskip18pt\vrule width1pt height1pt\hskip7pt\vrule width1pt
 height1pt\hskip7pt\vrule width1pt height1pt\hskip7pt\vrule width1pt
 height1pt\hfil}
\hbox{\hskip164pt\vrule width1pt height1pt\hskip7pt\vrule width1pt
 height1pt\hskip7pt\vrule width1pt height1pt\hskip7pt\vrule width1pt
 height1pt\hskip7pt\vrule width1pt height1pt\hskip7pt\vrule width1pt
 height1pt\hskip7pt\vrule width1pt height1pt\hskip7pt\vrule width1pt
 height1pt\hfil}
\vskip2pt
\hbox{\hskip22pt\vrule width1pt height1pt\hskip7pt\vrule width1pt
 height1pt\hskip7pt\vrule width1pt height1pt\hskip7pt\vrule width1pt
 height1pt\hfil}
\hbox{\hskip168pt\vrule width1pt height1pt\hskip7pt\vrule width1pt
 height1pt\hskip7pt\vrule width1pt height1pt\hskip7pt\vrule width1pt
 height1pt\hskip7pt\vrule width1pt height1pt\hskip7pt\vrule width1pt
 height1pt\hskip7pt\vrule width1pt height1pt\hskip7pt\vrule width1pt
 height1pt\hfil}
\vskip1pt
\hbox{\hskip196pt\vrule width1pt height1pt\hfil}
\hbox{\hskip26pt\vrule width1pt height1pt\hskip7pt\vrule width1pt
 height1pt\hskip7pt\vrule width1pt height1pt\hskip154pt\vrule width1pt
 height1pt\hfil}
\hbox{\hskip172pt\vrule width1pt height1pt\hskip7pt\vrule width1pt
 height1pt\hskip7pt\vrule width1pt height1pt\hskip7pt\vrule width1pt
 height1pt\hskip7pt\vrule width1pt height1pt\hskip7pt\vrule width1pt
 height1pt\hskip7pt\vrule width1pt height1pt\hfil}
\vskip2pt
\hbox{\hskip20pt\vrule width1pt height1pt\hskip1pt\vrule width1pt
 height1pt\hskip7pt\vrule width1pt height1pt\hskip7pt\vrule width1pt
 height1pt\hskip7pt\vrule width1pt height1pt\hskip6pt\vrule width1pt
 height1pt\hskip56pt\vrule width1pt height1pt\hskip55pt\vrule width1pt
 height1pt\hskip32pt\vrule width1pt height1pt\hskip23pt\vrule width1pt
 height1pt\hskip32pt\vrule width1pt height1pt\hfil}
\hbox{\hskip77pt\vrule width1pt height1pt\hskip89pt\vrule width2pt
 height1pt\hskip7pt\vrule width1pt height1pt\hskip7pt\vrule width1pt
 height1pt\hskip7pt\vrule width1pt height1pt\hskip7pt\vrule width1pt
 height1pt\hskip7pt\vrule width1pt height1pt\hskip7pt\vrule width1pt
 height1pt\hfil}
\hbox{\hskip76pt\vrule width1pt height1pt\hskip90pt\vrule width1pt
 height1pt\hfil}
\hbox{\hskip22pt\vrule width1pt height1pt\hskip32pt\vrule width1pt
 height1pt\hskip52pt\vrule width1pt height1pt\hskip59pt\vrule width1pt
 height1pt\hskip52pt\vrule width1pt height1pt\hskip32pt\vrule width1pt
 height1pt\hfil}
\hbox{\hskip23pt\vrule width1pt height1pt\hskip2pt\vrule width1pt
 height1pt\hskip7pt\vrule width1pt height1pt\hskip7pt\vrule width1pt
 height1pt\hskip7pt\vrule width1pt height1pt\hskip5pt\vrule width1pt
 height1pt\hskip50pt\vrule width1pt height1pt\hskip61pt\vrule width1pt
 height1pt\hskip50pt\vrule width1pt height1pt\hskip32pt\vrule width1pt
 height1pt\hfil}
\hbox{\hskip24pt\vrule width1pt height1pt\hskip32pt\vrule width1pt
 height1pt\hskip48pt\vrule width1pt height1pt\hskip63pt\vrule width1pt
 height1pt\hskip1pt\vrule width1pt height1pt\hskip7pt\vrule width1pt
 height1pt\hskip7pt\vrule width1pt height1pt\hskip7pt\vrule width1pt
 height1pt\hskip6pt\vrule width1pt height1pt\hskip8pt\vrule width1pt
 height1pt\hskip6pt\vrule width1pt height1pt\hskip32pt\vrule width1pt
 height1pt\hfil}
\hbox{\hskip25pt\vrule width1pt height1pt\hskip32pt\vrule width1pt
 height1pt\hskip13pt\vrule width1pt height1pt\hskip32pt\vrule width1pt
 height1pt\hskip65pt\vrule width1pt height1pt\hskip32pt\vrule width1pt
 height1pt\hskip13pt\vrule width1pt height1pt\hskip32pt\vrule width1pt
 height1pt\hfil}
\hbox{\hskip26pt\vrule width1pt height1pt\hskip32pt\vrule width1pt
 height1pt\hskip11pt\vrule width1pt height1pt\hskip32pt\vrule width1pt
 height1pt\hskip67pt\vrule width1pt height1pt\hskip32pt\vrule width1pt
 height1pt\hskip11pt\vrule width1pt height1pt\hskip32pt\vrule width1pt
 height1pt\hfil}
\hbox{\hskip27pt\vrule width2pt height1pt\hskip1pt\vrule width1pt
 height1pt\hskip7pt\vrule width1pt height1pt\hskip7pt\vrule width1pt
 height1pt\hskip7pt\vrule width1pt height1pt\hskip5pt\vrule width2pt
 height1pt\hskip40pt\vrule width2pt height1pt\hskip69pt\vrule width2pt
 height1pt\hskip40pt\vrule width2pt height1pt\hskip31pt\vrule width2pt
 height1pt\hfil}
\hbox{\hskip29pt\vrule width1pt height1pt\hskip32pt\vrule width1pt
 height1pt\hskip38pt\vrule width1pt height1pt\hskip73pt\vrule width2pt
 height1pt\hskip7pt\vrule width1pt height1pt\hskip7pt\vrule width1pt
 height1pt\hskip7pt\vrule width1pt height1pt\hskip7pt\vrule width1pt
 height1pt\hskip5pt\vrule width1pt height1pt\hskip32pt\vrule width1pt
 height1pt\hfil}
\hbox{\hskip30pt\vrule width2pt height1pt\hskip31pt\vrule width1pt
 height1pt\hskip35pt\vrule width2pt height1pt\hskip75pt\vrule width2pt
 height1pt\hskip35pt\vrule width1pt height1pt\hskip31pt\vrule width2pt
 height1pt\hfil}
\hbox{\hskip32pt\vrule width2pt height1pt\hskip30pt\vrule width3pt
 height1pt\hskip30pt\vrule width2pt height1pt\hskip79pt\vrule width2pt
 height1pt\hskip30pt\vrule width3pt height1pt\hskip30pt\vrule width2pt
 height1pt\hfil}
\hbox{\hskip34pt\vrule width2pt height1pt\hskip6pt\vrule width1pt
 height1pt\hskip7pt\vrule width1pt height1pt\hskip7pt\vrule width1pt
 height1pt\hskip3pt\vrule width2pt height1pt\hskip3pt\vrule width2pt
 height1pt\hskip26pt\vrule width2pt height1pt\hskip83pt\vrule width2pt
 height1pt\hskip26pt\vrule width2pt height1pt\hskip3pt\vrule width2pt
 height1pt\hskip26pt\vrule width2pt height1pt\hfil}
\hbox{\hskip36pt\vrule width3pt height1pt\hskip20pt\vrule width3pt
 height1pt\hskip7pt\vrule width3pt height1pt\hskip20pt\vrule width3pt
 height1pt\hskip87pt\vrule width3pt height1pt\hskip11pt\vrule width1pt
 height1pt\hskip7pt\vrule width4pt height1pt\hskip7pt\vrule width3pt
 height1pt\hskip20pt\vrule width3pt height1pt\hfil}
\hbox{\hskip39pt\vrule width4pt height1pt\hskip12pt\vrule width4pt
 height1pt\hskip13pt\vrule width4pt height1pt\hskip12pt\vrule width4pt
 height1pt\hskip93pt\vrule width3pt height1pt\hskip13pt\vrule width4pt
 height1pt\hskip13pt\vrule width4pt height1pt\hskip12pt\vrule width4pt
 height1pt\hfil}
\hbox{\hskip43pt\vrule width12pt height1pt\hskip21pt\vrule width12pt
 height1pt\hskip101pt\vrule width12pt height1pt\hskip21pt\vrule width
12pt height1pt\hfil}
\vfil}
%
%
\vskip -208pt
\vbox to 208pt{\vfil\hbox{\unskip
\hskip13pt\vbox{\hrule height2pt width1pt\vskip110pt\hrule height2pt
 width1pt\vskip46pt}\unskip
\vbox{\hrule height13pt width1pt\vskip100pt\hrule height12pt width1pt
\vskip41pt}\unskip
\vbox{\hrule height5pt width1pt\vskip2pt\hrule height5pt width1pt
\vskip4pt\hrule height5pt width1pt\vskip92pt\hrule height4pt width1pt
\vskip3pt\hrule height4pt width1pt\vskip5pt\hrule height4pt width1pt
\vskip37pt}\unskip
\vbox{\hrule height3pt width1pt\vskip7pt\hrule height3pt width1pt
\vskip10pt\hrule height3pt width1pt\vskip87pt\hrule height3pt width1pt
\vskip20pt\hrule height3pt width1pt\vskip34pt}\unskip
\vbox{\hrule height2pt width1pt\vskip10pt\hrule height2pt width1pt
\vskip14pt\hrule height2pt width1pt\vskip83pt\hrule height2pt width1pt
\vskip26pt\hrule height2pt width1pt\vskip32pt}\unskip
\vbox{\hrule height2pt width1pt\vskip30pt\hrule height2pt width1pt
\vskip79pt\hrule height2pt width1pt\vskip30pt\hrule height2pt width1pt
\vskip30pt}\unskip
\vbox{\hrule height2pt width1pt\vskip34pt\hrule height2pt width1pt
\vskip75pt\hrule height2pt width1pt\vskip34pt\hrule height2pt width1pt
\vskip28pt}\unskip
\hskip1pt\vbox{\hrule height2pt width1pt\vskip111pt\hrule height2pt
 width1pt\vskip40pt\hrule height2pt width1pt\vskip25pt}\unskip
\hskip6pt\vbox{\hrule height2pt width1pt\vskip110pt\hrule height2pt
 width1pt\vskip43pt}\unskip
\hskip1pt\vbox{\hrule height2pt width1pt\vskip110pt\hrule height2pt
 width1pt\vskip46pt}\unskip
\hskip1pt\vbox{\hrule height2pt width1pt\vskip110pt\hrule height2pt
 width1pt\vskip43pt}\unskip
\hskip4pt\vbox{\hrule height7pt width1pt\vskip105pt\hrule height7pt
 width1pt\vskip40pt}\unskip
\hskip4pt\vbox{\hrule height2pt width1pt\vskip126pt}\unskip
\hskip4pt\vbox{\hrule height3pt width1pt\vskip3pt\hrule height2pt
 width1pt\vskip106pt\hrule height2pt width1pt\vskip4pt\hrule height4pt
 width1pt\vskip41pt}\unskip
\vbox{\hrule height2pt width1pt\vskip15pt\hrule height3pt width1pt
\vskip92pt\hrule height3pt width1pt\vskip14pt\hrule height4pt width1pt
\vskip37pt}\unskip
\vbox{\hrule height3pt width1pt\vskip34pt}\unskip
\vbox{\hrule height2pt width1pt\vskip27pt\hrule height2pt width1pt
\vskip111pt\hrule height3pt width1pt\vskip31pt}\unskip
\vbox{\hrule height2pt width1pt\vskip30pt}\unskip
\vbox{\hrule height2pt width1pt\vskip28pt}\unskip
\hskip1pt\vbox{\hrule height2pt width1pt\vskip25pt}\unskip
\hskip21pt\vbox{\hrule height2pt width1pt\vskip40pt\hrule height2pt
 width1pt\vskip69pt\hrule height2pt width1pt\vskip67pt}\unskip
\hskip1pt\vbox{\hrule height2pt width1pt\vskip34pt\hrule height2pt
 width1pt\vskip75pt\hrule height2pt width1pt\vskip64pt}\unskip
\vbox{\hrule height2pt width1pt\vskip30pt\hrule height2pt width1pt
\vskip79pt\hrule height2pt width1pt\vskip62pt}\unskip
\vbox{\hrule height2pt width1pt\vskip26pt\hrule height2pt width1pt
\vskip83pt\hrule height2pt width1pt\vskip27pt\hrule height2pt width1pt
\vskip31pt}\unskip
\vbox{\hrule height3pt width1pt\vskip20pt\hrule height3pt width1pt
\vskip87pt\hrule height3pt width1pt\vskip57pt}\unskip
\vbox{\hrule height4pt width1pt\vskip12pt\hrule height4pt width1pt
\vskip93pt\hrule height4pt width1pt\vskip14pt\hrule height2pt width1pt
\vskip37pt}\unskip
\vbox{\hrule height12pt width1pt\vskip101pt\hrule height6pt width1pt
\vskip3pt\hrule height2pt width1pt\vskip42pt}\unskip
\hskip10pt\vbox{\hrule height2pt width1pt\vskip111pt\hrule height2pt
 width1pt\vskip43pt}\unskip
\hskip1pt\vbox{\hrule height2pt width1pt\vskip111pt\hrule height2pt
 width1pt\vskip46pt}\unskip
\hskip1pt\vbox{\hrule height2pt width1pt\vskip111pt\hrule height2pt
 width1pt\vskip43pt}\unskip
\hskip4pt\vbox{\hrule height2pt width1pt\vskip111pt\hrule height2pt
 width1pt\vskip40pt}\unskip
\vbox{\hrule height2pt width1pt\vskip111pt\hrule height2pt width1pt
\vskip42pt}\unskip
\vbox{\hrule height2pt width1pt\vskip111pt\hrule height2pt width1pt
\vskip44pt}\unskip
\hskip3pt\vbox{\hrule height2pt width1pt\vskip40pt\hrule height2pt
 width1pt\vskip69pt\hrule height2pt width1pt\vskip40pt\hrule height2pt
 width1pt\vskip25pt}\unskip
\hskip1pt\vbox{\hrule height2pt width1pt\vskip34pt\hrule height2pt
 width1pt\vskip75pt\hrule height2pt width1pt\vskip34pt\hrule height2pt
 width1pt\vskip28pt}\unskip
\vbox{\hrule height2pt width1pt\vskip30pt\hrule height2pt width1pt
\vskip79pt\hrule height2pt width1pt\vskip30pt\hrule height2pt width1pt
\vskip30pt}\unskip
\vbox{\hrule height2pt width1pt\vskip26pt\hrule height2pt width1pt
\vskip83pt\hrule height2pt width1pt\vskip26pt\hrule height2pt width1pt
\vskip32pt}\unskip
\vbox{\hrule height3pt width1pt\vskip11pt\hrule height2pt width1pt
\vskip7pt\hrule height3pt width1pt\vskip87pt\hrule height3pt width1pt
\vskip11pt\hrule height2pt width1pt\vskip7pt\hrule height3pt width1pt
\vskip34pt}\unskip
\vbox{\hrule height4pt width1pt\vskip5pt\hrule height4pt width1pt
\vskip3pt\hrule height4pt width1pt\vskip93pt\hrule height4pt width1pt
\vskip5pt\hrule height4pt width1pt\vskip3pt\hrule height4pt width1pt
\vskip37pt}\unskip
\vbox{\hrule height12pt width1pt\vskip101pt\hrule height12pt width1pt
\vskip41pt}\unskip
\vbox{\hrule height4pt width1pt\vskip109pt\hrule height4pt width1pt
\vskip44pt}\unskip
\vbox{\hrule height2pt width1pt\vskip111pt\hrule height2pt width1pt
\vskip44pt}\unskip
\hskip40pt\vbox{\hrule height2pt width1pt\vskip111pt\hrule height2pt
 width1pt\vskip46pt}\unskip
\vbox{\hrule height12pt width1pt\vskip101pt\hrule height12pt width1pt
\vskip41pt}\unskip
\vbox{\hrule height4pt width1pt\vskip3pt\hrule height4pt width1pt
\vskip5pt\hrule height4pt width1pt\vskip93pt\hrule height4pt width1pt
\vskip3pt\hrule height4pt width1pt\vskip5pt\hrule height4pt width1pt
\vskip37pt}\unskip
\vbox{\hrule height3pt width1pt\vskip20pt\hrule height3pt width1pt
\vskip87pt\hrule height3pt width1pt\vskip20pt\hrule height3pt width1pt
\vskip34pt}\unskip
\vbox{\hrule height2pt width1pt\vskip26pt\hrule height2pt width1pt
\vskip83pt\hrule height2pt width1pt\vskip26pt\hrule height2pt width1pt
\vskip32pt}\unskip
\vbox{\hrule height2pt width1pt\vskip30pt\hrule height2pt width1pt
\vskip79pt\hrule height2pt width1pt\vskip30pt\hrule height2pt width1pt
\vskip30pt}\unskip
\vbox{\hrule height2pt width1pt\vskip34pt\hrule height2pt width1pt
\vskip75pt\hrule height2pt width1pt\vskip34pt\hrule height2pt width1pt
\vskip28pt}\unskip
\hskip1pt\vbox{\hrule height2pt width1pt\vskip40pt\hrule height2pt
 width1pt\vskip69pt\hrule height2pt width1pt\vskip67pt}\unskip
\hskip2pt\vbox{\hrule height2pt width1pt\vskip183pt}\unskip
\hskip3pt\vbox{\hrule height2pt width1pt\vskip157pt}\unskip
\vbox{\hrule height2pt width1pt\vskip43pt}\unskip
\vbox{\hrule height2pt width1pt\vskip160pt}\unskip
\vbox{\hrule height2pt width1pt\vskip46pt}\unskip
\vbox{\hrule height3pt width1pt\vskip157pt}\unskip
\vbox{\hrule height2pt width1pt\vskip43pt}\unskip
\hskip3pt\vbox{\hrule height7pt width1pt\vskip154pt}\unskip
\vbox{\hrule height7pt width1pt\vskip40pt}\unskip
\hskip1pt\vbox{\hrule height2pt width1pt\vskip126pt}\unskip
\hskip1pt\vbox{\hrule height2pt width1pt\vskip13pt}\unskip
\hskip4pt\vbox{\hrule height12pt width1pt\vskip101pt\hrule height4pt
 width1pt\vskip4pt\hrule height3pt width1pt\vskip42pt}\unskip
\vbox{\hrule height4pt width1pt\vskip12pt\hrule height4pt width1pt
\vskip93pt\hrule height4pt width1pt\vskip14pt\hrule height3pt width1pt
\vskip36pt}\unskip
\vbox{\hrule height3pt width1pt\vskip20pt\hrule height3pt width1pt
\vskip87pt\hrule height3pt width1pt\vskip57pt}\unskip
\vbox{\hrule height2pt width1pt\vskip26pt\hrule height2pt width1pt
\vskip82pt\hrule height3pt width1pt\vskip60pt}\unskip
\vbox{\hrule height2pt width1pt\vskip30pt\hrule height2pt width1pt
\vskip79pt\hrule height2pt width1pt\vskip62pt}\unskip
\vbox{\hrule height3pt width1pt\vskip34pt\hrule height2pt width1pt
\vskip75pt\hrule height2pt width1pt\vskip64pt}\unskip
\hskip1pt\vbox{\hrule height2pt width1pt\vskip40pt\hrule height2pt
 width1pt\vskip69pt\hrule height2pt width1pt\vskip67pt}\unskip
\hskip1pt\vbox{\hrule height2pt width1pt\vskip7pt\hrule height2pt
 width1pt\vskip126pt}\unskip
\hskip19pt\vbox{\hrule height2pt width1pt\vskip25pt}\unskip
\hskip1pt\vbox{\hrule height2pt width1pt\vskip28pt}\unskip
\vbox{\hrule height2pt width1pt\vskip30pt}\unskip
\vbox{\hrule height2pt width1pt\vskip110pt\hrule height2pt width1pt
\vskip27pt\hrule height2pt width1pt\vskip32pt}\unskip
\vbox{\hrule height3pt width1pt\vskip34pt}\unskip
\vbox{\hrule height2pt width1pt\vskip15pt\hrule height2pt width1pt
\vskip93pt\hrule height2pt width1pt\vskip14pt\hrule height4pt width1pt
\vskip37pt}\unskip
\vbox{\hrule height2pt width1pt\vskip3pt\hrule height2pt width1pt
\vskip104pt\hrule height2pt width1pt\vskip3pt\hrule height6pt width1pt
\vskip41pt}\unskip
\hskip10pt\vbox{\hrule height2pt width1pt\vskip111pt\hrule height2pt
 width1pt\vskip43pt}\unskip
\hskip1pt\vbox{\hrule height2pt width1pt\vskip111pt\hrule height2pt
 width1pt\vskip46pt}\unskip
\hskip1pt\vbox{\hrule height2pt width1pt\vskip111pt\hrule height2pt
 width1pt\vskip43pt}\unskip
\hskip4pt\vbox{\hrule height2pt width1pt\vskip111pt\hrule height2pt
 width1pt\vskip40pt}\unskip
\vbox{\hrule height2pt width1pt\vskip111pt\hrule height2pt width1pt
\vskip42pt}\unskip
\vbox{\hrule height2pt width1pt\vskip111pt\hrule height2pt width1pt
\vskip44pt}\unskip
\hskip3pt\vbox{\hrule height2pt width1pt\vskip40pt\hrule height2pt
 width1pt\vskip69pt\hrule height2pt width1pt\vskip40pt\hrule height2pt
 width1pt\vskip25pt}\unskip
\hskip1pt\vbox{\hrule height2pt width1pt\vskip34pt\hrule height2pt
 width1pt\vskip75pt\hrule height2pt width1pt\vskip34pt\hrule height2pt
 width1pt\vskip28pt}\unskip
\vbox{\hrule height2pt width1pt\vskip30pt\hrule height2pt width1pt
\vskip79pt\hrule height2pt width1pt\vskip30pt\hrule height2pt width1pt
\vskip30pt}\unskip
\vbox{\hrule height2pt width1pt\vskip26pt\hrule height2pt width1pt
\vskip83pt\hrule height2pt width1pt\vskip26pt\hrule height2pt width1pt
\vskip32pt}\unskip
\vbox{\hrule height3pt width1pt\vskip11pt\hrule height2pt width1pt
\vskip7pt\hrule height3pt width1pt\vskip87pt\hrule height3pt width1pt
\vskip11pt\hrule height2pt width1pt\vskip7pt\hrule height3pt width1pt
\vskip34pt}\unskip
\vbox{\hrule height4pt width1pt\vskip5pt\hrule height4pt width1pt
\vskip3pt\hrule height4pt width1pt\vskip93pt\hrule height4pt width1pt
\vskip5pt\hrule height4pt width1pt\vskip3pt\hrule height4pt width1pt
\vskip37pt}\unskip
\vbox{\hrule height12pt width1pt\vskip101pt\hrule height12pt width1pt
\vskip41pt}\unskip
\vbox{\hrule height4pt width1pt\vskip109pt\hrule height4pt width1pt
\vskip44pt}\unskip
\vbox{\hrule height2pt width1pt\vskip111pt\hrule height2pt width1pt
\vskip44pt}\unskip
\hfil}}}
\caption{Different overlaps of the surfaces $\Sigma_1$ and $\Sigma_2$
on a plane, determined by the possible liftings of the loops
$\partial\Sigma_1$ and $\partial\Sigma_2$ away from the plane.}
\end{figure}
}

For non intersecting loops this is a well defined, non singular
object in the Chern-Simons theory, which depends only on the
linking of the two loops $\Gamma_1$ and $\Gamma_2$~\cite{witten}.
Assuming this to be non-trivial (and non-singular), the two loops
cannot be flat and lying on the same plane, so the previous
construction fails.  But once the  ratio~(\ref{two-loops}) has
been computed in the Chern-Simons theory, we can take the limit
in which the two loops collapse onto a single plane.  This is a
singular limit in which the loops necessarily intersect each other.
Their linking is not well defined any more, and the value
of~(\ref{two-loops}) depends on the initial non-singular loops
used in the computation.  However, at the classical level, before
the functional integral is performed, we can repeat the previous
construction with no difficulties for any arrangement of loops on
the plane~\cite{bralic}.  Thus, formally we can write
\beq
\frac{\langle \J[\Sigma_1\cup\Sigma_2] \rangle_{ferm}}
     {\langle \J[\Sigma_1] \rangle_{ferm}
      \langle \J[\Sigma_2] \rangle_{ferm}} =
\frac{\langle W[\partial\Sigma_1] W[\partial\Sigma_2] \rangle_{CS}}
     {\langle W[\partial\Sigma_1] \rangle_{CS}
      \langle W[\partial\Sigma_2] \rangle_{CS}}
\label{two-surf}
\eeq
where both surfaces $\Sigma_1$ and $\Sigma_2$ are contained in the
same plane.  As we just stated, the bosonic side of this relation
will be ill defined in general.  But it can be given a well defined
meaning by lifting the loops $\partial\Sigma_i$ from the plane to non
intersecting three-dimensional loops $\Gamma_i$.  This can be done
in different ways, specifying different linkings of the loops
$\Gamma_i$ compatible with the intersections of their projections
$\partial\Sigma_i$ onto the plane.  Correspondingly, in the fermionic side,
the surface $\Sigma_1\cup\Sigma_2$ must be complemented with a
prescription stating the way in which the two surfaces $\Sigma_i$
overlap.  The different possible liftings of the loops specify
different overlaps of the surfaces, as ilustrated in Fig.~(1).
In this way, relation~(\ref{two-surf}) (and its generalizations)
can be viewed as a defining relation, through bosonization, of
the vaccuum expectation value of the flux of the fermionic current
through surfaces with foldings.

\section{Bosonization in $d=4$}
In ref.\cite{5} we have applied our bosonization approach to the study
of vector and axial-vector currents in $d=4$ dimensions. I will
briefly describe the main results of this work.
Consider the
generating functional for a massless  Dirac
field in  $3+1$ (Euclidean) dimensions, coupled to Abelian vector ($s_\mu$)
and axial-vector ($t_\mu$) external sources
$$ Z ( s_\mu , t_\mu ) \;=\; \int \D {\bar \psi} \D \psi \,
\exp \left[ - S( {\bar \psi} , \psi ; s_\mu , t_\mu ) \right]$$
\beq
S( {\bar \psi} , \psi ; s_\mu , t_\mu ) \;=\; -i  \int d^4 x \; {\bar \psi}
\, (\id  - \as  -  \gamma_5 \bs ) \, \psi
\label{strt}
\eeq
where we have adopted for the $\gamma$-matrices the following conventions:
\beq
\gamma_\mu^\dagger \;=\; \gamma_\mu \;\;,\;\;
\gamma_5^\dagger \;=\; \gamma_5 \;\;,\;\;
\{ \gamma_\mu \,,\,\gamma_\nu \} \,=\, 2 \, \delta_{\mu\nu} \;.
\eeq
The addition of the source $s_\mu$ is  due to the fact that,
in four dimensions, the vector and axial currents are {\em independent \/}
fermionic bilinears. Thus not all the information provided by
$Z (s_\mu, t_\mu)$ can be obtained from, say, $Z (s_\mu , 0)$.
In two dimensions, because of the smaller number of generators for
the Dirac algebra, these two currents are related, and the bosonization
rule for one of the currents also implies the proper rule for the other.

We now consider the following change of variables

\beq
\psi (x) \,=\, e^{i \theta (x) - i \gamma_5 \alpha (x) } \psi' (x) \;,\;
{\bar \psi} (x)\,=\,{\bar \psi}' (x) e^{- i \theta (x) - i \gamma_5 \alpha(x)}
\;.
\label{trafo}
\eeq
In terms of the new variables, the generating functional (\ref{strt}) reads
\beq
\Z ( s_\mu , t_\mu ) \;=\; \int \D {\bar \psi} \D \psi \,
J (\alpha ; s_\mu , t_\mu) \,
\exp \left[ - S( {\bar \psi} , \psi \;;\; s_\mu + \partial_\mu
\theta , t_\mu + \partial_\mu \alpha) \right]
\label{scnd}
\eeq
where the primed fermionic fields have been renamed as unprimed for
the sake of simplicity, and $J$ is the anomalous Jacobian corresponding to
this fermionic change of variables, a well-known consequence of the chiral
anomaly~\cite{jaco}. This Jacobian is evaluated by using the standard
Fujikawa's recipe accomodated so as to get the consistent anomaly.
The answer is  \cite{5}
\beq
J (\alpha ; s_\mu , t_\mu)\;=\; \exp \left[ \frac{1}{4 \pi^2} \int
d^4 x \, \alpha (x) \, \epsilon_{\mu\nu\rho\sigma}
(\partial_\mu s_\nu \partial_\rho s_\sigma \,+\, \frac{1}{3}
\partial_\mu t_\nu \partial_\rho t_\sigma)\right] \;.
\label{jaco}
\eeq
Now, as (\ref{trafo}) is a change of variables, $Z$ cannot depend on either
$\theta$ or $\alpha$. Thus $\theta$ and $\alpha$ can be integrated out
without other effect than the introduction of  an irrelevant constant
factor in $Z$, which we ignore. Integration over $\theta$ and $\alpha$ is
equivalent to integration over two flat Abelian vector fields
$\theta_\mu$ and $\alpha_\mu$:

$$\theta_\mu \;=\; \partial_\mu \theta \;\;\; ,
\;\;\; \alpha_\mu \;=\; \partial_\mu \alpha$$
\beq
f_{\mu\nu}(\theta) \; \equiv \; \partial_\mu \theta_\nu - \partial_\nu
\theta_\mu \,=\,0
                           \;\;,\;\;
f_{\mu\nu}(\alpha) \; \equiv \; \partial_\mu \alpha_\nu - \partial_\nu
\alpha_\mu \,=\,0 \;.
\label{defs}
\eeq
Writing
\beq
J(\alpha_\mu ; s_\mu , t_\mu)\;=\; \exp \left[ - \frac{1}{4 \pi^2} \int
d^4 x \, \alpha_\mu (x) \, \epsilon_{\mu\nu\rho\sigma}
(s_\nu \partial_\rho s_\sigma \,+\, \frac{1}{3}
t_\nu \partial_\rho t_\sigma)\right] \;.
\label{jaco1}
\eeq
eq.(\ref{scnd}) becomes
$$
\Z \;=\; \int \D \theta_\mu \, \D \alpha_\mu \,\D {\bar \psi}\,\D \psi \;
\delta [f_{\mu\nu}(\theta)] \,
\delta [f_{\mu\nu} (\alpha)]\;
J(\alpha_\mu ; s_\mu , t_\mu)
$$
\beq
\exp \left[ - S( {\bar \psi} , \psi ; s_\mu + \theta_\mu , t_\mu +
\alpha_\mu ) \right] \;.
\label{trd}
\eeq
Formally integrating out the fermionic fields and making
the shift of variables
\beq
\theta_\mu \; \to \; \theta_\mu \,-\, s_\mu \;\;\;,\;\;\;
\alpha_\mu \; \to \; \alpha_\mu \,-\, t_\mu \; ,
\eeq
(\ref{trd}) leads to
$$
Z (s_\mu,t_\mu)\,=\, \int \D \theta_\mu \, \D \alpha_\mu
\,\delta[f_{\mu\nu}(\theta
-s)] \delta[f_{\mu\nu}(\alpha-t)]\, J(\alpha_\mu - t_\mu  ; s_\mu , t_\mu)
$$
\beq
\times \det ( \not \! \partial + i \not \! \theta + i \gamma_5 \not
\! \alpha) \;.
\label{frth}
\eeq
As before, we  exponentiate the
two functional delta functions in (\ref{frth}) using two antisymmetric tensor
fields
$A_{\mu\nu}$ and $B_{\mu\nu}$ as Lagrange multipliers
\newpage
$$ Z (s_\mu, t_\mu) \;=\; \int \D A_{\mu\nu} \, \D B_{\mu\nu} \,
\D \theta_\mu \, \D \alpha_\mu \;\; J(\alpha_\mu - t_\mu  ; s_\mu , t_\mu)$$
$$ \times \exp\left({i \int d^4 x [\epsilon_{\mu\nu\rho\sigma}
A_{\mu\nu} (\partial_\rho \theta_\sigma - \partial_\rho s_\sigma) \;+\;
\epsilon_{\mu\nu\rho\sigma} B_{\mu\nu}  ( \partial_\rho \alpha_\sigma -
\partial_\rho t_\sigma ) ]}\right) \;$$
\beq
\times \;\; \det ( \not \! \partial + i \not \! \theta +
i \gamma_5 \not \! \alpha) \;.
\label{fvth}
\eeq
The bosonized form of $Z$ can then be obtained by integrating out
$\theta_\mu$ and $\alpha_\mu$ in (\ref{fvth}). This produces
a generating functional with the tensor fields $A_{\mu\nu}$ and
$B_{\mu\nu}$ as dynamical variables. This step requires
the evaluation of the fermionic determinant, which of course
is necessarily non-exact.

At this stage we can already derive the rules that
map the vector and axial-vector currents into functions of the
bosonic fields $A_{\mu\nu}$ and $B_{\mu\nu}$. This
correspondence requires no approximation and may well be called `exact'.
These rules follow from elementary functional differentiation
\beq
j_{\mu} = \langle {\bar \psi} \gamma_\mu \psi \rangle
= -i \frac{\delta}{\delta s_\mu} \log \Z |_{s_\mu = 0}
= - \epsilon_{\mu\nu\rho\sigma} \partial_\nu A_{\rho\sigma}
\label{ve}
\eeq
\beq
j_{5 \mu} = \langle {\bar \psi} \gamma_5 \gamma_\mu \psi \rangle
= -i \frac{\delta}{\delta t_\mu} \log \Z |_{t_\mu = 0}
= - \epsilon_{\mu\nu\rho\sigma} \partial_\nu B_{\rho\sigma}
- \frac{i}{4\pi^2} \, \epsilon_{\mu\nu\rho\sigma}
s_\nu \partial_\rho s_\sigma \;.
\label{ax}
\eeq
 From the antisymmetry of the tensors $A_{\mu\nu}$ and $B_{\mu\nu}$,
we are entitled to derive the equations for the divergencies of the
currents:
\begin{eqnarray}
\partial_\mu j_\mu &=& 0 \nonumber\\
\partial_\mu j^5_\mu &=& - \frac{i}{8 \pi^2} \,
{\tilde F}_{\mu\nu}(s) F_{\mu\nu}(s) \,.
\end{eqnarray}
with
${\tilde F}_{\mu\nu} =(1/2) \epsilon_{\mu\nu\alpha\beta}F_{\alpha\beta}$.
We then see that the bosonization rule (\ref{ax}) correctly reproduces
the axial anomaly.

As before, although the bosonization recipe (\ref{ve})-(\ref{ax})
for associating the fermionic currents with expressions written
in terms of bosonic fields is exact, the bosonic action governing
the boson field dynamics cannot be evaluated in an exact form in
$d>2$ dimensions. Different approximations for computing the
fermionic determinant would yield alternative effective
bosonic actions valid in different regimes.
In ref.\cite{5}  the fermionic determinant
in (\ref{fvth}) was evaluated to second order in the fields $\theta_\mu$ and
$\alpha_\mu$. The use of this quadratic approximation can be
motivated by the same kind of arguments (see in particular the
`quasi-theorem') used in ref.~\cite{Fro}. Calling
\beq
\det (\not \! \partial \, + \, i \not \! \theta \,+\, i \gamma_5
\not \! \alpha) \;=\; \exp \left[ W (\theta_\mu , \alpha_\mu) \right] \; .
\label{defw}
\eeq
the answer for the renormalized  $W$ is
$$
W (\theta_\mu , \alpha_\mu) \;=\;-\frac{1}{2} \, \int d^4 x d^4 y \,
\left[ \theta_\mu (x) \delta^\perp_{\mu\nu} F(x-y) \theta_\nu (y)
\right.
$$
\beq
\left. +\,\alpha_\mu (x) \delta^\perp_{\mu\nu} G(x-y) \alpha_\nu (y)
\,+\, m^2 \, \alpha_\mu (x) \delta^\parallel_{\mu\nu} \delta (x-y)
\alpha_\nu (y) \right]
\label{wren}
\eeq
where
\beq
G(x-y) \;=\; F(x-y) \,+\, m^2 \delta (x-y) \;.
\eeq
With this result, one easily finds for the generating
functional
\begin{eqnarray}
& & Z (s_\mu,t_\mu) \,=\, \exp [{\cal C} (s_\mu,t_\mu)] \,
\int \D A_{\mu\nu} \, \D B_{\mu\nu} \times \nonumber\\
& & \exp \left\{-i \int d^4 x [ s_\mu \epsilon_{\mu\nu\rho\sigma}
\partial_\nu A_{\rho\sigma} + t_\mu (\epsilon_{\mu\nu\rho\sigma}
\partial_\nu B_{\rho\sigma}+ \frac{i}{4 \pi^2}\epsilon_{\mu\nu\rho\sigma}
s_\nu \partial_\rho s_\sigma)] \right\} \times \nonumber\\
& &
\exp \left\{ - \frac{1}{3} \, \int d^4 x d^4 y [ A_{\mu\nu\rho}(x)
F^{-1}(x-y) A_{\mu\nu\rho}(y) + \right.\nonumber \\
& & \left. B_{\mu\nu\rho}(x)
G^{-1}(x-y) B_{\mu\nu\rho}(y) ] \right\} \times
\exp \left\{ - \frac{i}{4 \pi^2} \int d^4 x d^4 y
\partial_\mu B_{\nu\rho}(x) \times \right. \nonumber\\
& & \left. G^{-1}(x-y) \delta_{\mu\nu\rho ,
\alpha\beta\gamma} (s_\alpha \partial_\beta s_\gamma + \frac{1}{3}
t_\alpha \partial_\beta t_\gamma) \right\}
\label{boson}
\end{eqnarray}
where
\begin{eqnarray}
A_{\mu\nu\rho} &=& \partial_\mu A_{\nu\rho} + \partial_\nu A_{\rho\mu} +
\partial_\rho A_{\mu\nu} \nonumber\\
B_{\mu\nu\rho} &=& \partial_\mu B_{\nu\rho} + \partial_\nu B_{\rho\mu} +
\partial_\rho B_{\mu\nu} \nonumber\\
\delta_{\mu\nu\rho ,\alpha\beta\gamma}&=&
\det \left( \begin{array}{ccc}
\delta_{\mu\alpha} & \delta_{\mu\beta} & \delta_{\mu\gamma}\\
\delta_{\nu\alpha} & \delta_{\nu\beta} & \delta_{\nu\gamma}\\
\delta_{\rho\alpha} & \delta_{\rho\beta} & \delta_{\rho\gamma}
\end{array}
\right)
\end{eqnarray}
and
$$
{\cal C} (s_\mu,t_\mu) \,=\, \frac{1}{2 (2 \pi)^4} \int d^4 x d^4 y \,\left\{
[ s_\mu (x) \partial_\nu s_\lambda (x)\,+\, \frac{1}{3} t_\mu (x)\partial_\nu
t_\lambda (x) ] \right.
$$
$$
\delta_{\mu\nu\rho ,\alpha\beta\gamma} G^{-1} (x-y)
[ s_\alpha (y) \partial_\beta s_\gamma (y)\,+\, \frac{1}{3} t_\alpha
(y)\partial_\beta
t_\gamma (y) ]
$$
$$
+ \frac{1}{2 (2\pi)^4} \int d^4 x d^4 y \, {\cal G} (x)
\partial^{-2} G^{-1} (x-y) {\cal G} (y)
$$
\beq
\left. + \frac{1}{2 m^2 (2 \pi)^4} \int d^4 x d^4 y \,{\cal G}(x)
\partial^{-2} (x-y) {\cal G} (y)\right\}
\label{chu}
\eeq
where ${\cal G} = \epsilon_{\mu\nu\rho\lambda}( \partial_\mu  s_\nu
\partial_\rho
s_\lambda \,+\, \frac{1}{3} \partial_\mu t_\nu \partial_\rho t_\lambda)$.

In conclusion, we have applied our bosonization technique
to the case of massless Dirac fermions in
four dimensions in the presence of both vector and axial-vector
sources. This has allowed us to find the bosonization rules
for both fermionic currents, eqs.(\ref{ve})-(\ref{ax}),
in terms of Kalb-Ramond bosonic
fields. While the bosonization rule for the vector current
can be written in a natural and compact form, reminiscent
of the well-known two-dimensional bosonization rule,
\beq
\bar \psi \gamma_\mu \psi \to
-\epsilon_{\mu \nu \rho \sigma} \partial_\nu A_{\rho \sigma} \, ,
\label{ult}
\eeq
the result for the axial current is more involved and includes
the vector source
\beq
{\bar \psi} \gamma_5 \gamma_\mu \psi \to
- \epsilon_{\mu\nu\rho\sigma} \partial_\nu B_{\rho\sigma}
- \frac{i}{4\pi^2} \, \epsilon_{\mu\nu\rho\sigma}
s_\nu \partial_\rho s_\sigma \;.
\label{ultult}
\eeq
In our approach,
this is a consequence of the anomalous
behaviour of the fermionic measure under axial gauge transformations and
in this way
the bosonic form of the axial current correctly yields its
anomalous divergence.
We want to comment on the possibility of considering the particular case
of a purely chiral external source ($s_\mu \equiv \pm t_\mu$), and obtaining
a bosonized version for this model. The Kalb-Ramond field then
corresponds to a particular `chiral' combination of $A$ and $B$,
namely $A_{\mu\nu} \pm B_{\mu\nu}$.

\newpage
\vspace{1 cm}

\underline{Acknowledgements}:  F.A.S.
is partially suported
by CONICET and CICBA, Argentina and a
Commission of the European Communities
contract No:C11*-CT93-0315.


\begin{thebibliography}{99}
%
\bibitem{1} E.~Fradkin and F.A.~Schaposnik,
 {\it The Fermion-Boson Mapping in Three Dimensional
 Quantum Field Theory},
Phys. Lett. {\bf B338} (1994) 253.
\bibitem{2} N.~Brali\'c, E.~Fradkin, M.V.~Man\'\i as and F.A.~Schaposnik,
{\it Bosonization of Three Dimensional
Non-Abelian Fermionic Field Theories},
Nucl. Phys. {\bf B446} (1995) 144.
\bibitem{3} F.A.~Schaposnik,
{\it A Comment on Bosonization in $d \geq 2$ dimensions},
Phys. Lett. {\bf B356} (1995) 39.
\bibitem{4} J.C.~Le Guillou, C.~N\'u\~nez and  F.A.~Schaposnik,
{\it Current Algebra and Bosonization in Three Dimensions},
Ann. of Phys. (N.Y.) {\bf 251} (1996) 426.
\bibitem{5} C.D.~Fosco and F.A.~Schaposnik,
{\it Bosonization of Vector and Axial-Vector currents in
$3+1$ dimensions},
Phys. Lett. {\bf B391} (1997) 136.
\bibitem{6} J.C.~Le Guillou, E.~Moreno, C.~N\'u\~nez and
 F.A.~Schaposnik, {\it Non-abelian bosonization in two and three
dimensions},
Nucl. Phys. {\bf B484} (1997) 682.
\bibitem{7} J.C.~Le Guillou, E.~Moreno, C.~N\'u\~nez and
F.A.~Schaposnik, {\it On three dimensional bosonization},
hep-th 9703048, unpublished.
\bibitem{coleman} S.Coleman, Phys. Rev. {\bf D11} (1975) 2088;
S. Mandelstam, Phys. Rev. {\bf D11} (1975) 3026.
\bibitem{mattis} E. Lieb and D. Mattis, J. Math.
Phys. {\bf 6} (1965) 304; A. Luther and I. Peschel, Phys. Rev.
{\bf B12} (1975) 3908.
%
\bibitem{lut} A.~Luther, Phys. Rev. {\bf D19} (1979) 320.
\bibitem{pol} A.M.~Polyakov, Mod. Phys. Lett. {\bf A3} (1988) 325.
\bibitem{lusch} M.~Luscher, Nucl. Phys. {\bf B326} (1989) 557.
\bibitem{hal} F.D.M. Haldane, Helv. Phys. Acta {\bf 65} (1992) 52.
\bibitem{kov} A. Kovner and P.S. Kurzepa, Phys. Lett. {\bf B321} (1994) 129.
\bibitem{Mar}E.C.~Marino, {Phys. Lett.} {\bf B263} (1991), 63.
\bibitem{Fro}  J. Frohlich, R. G\"otschmann and P.A. Marchetti,
J.Phys. {\bf A28} (1995) 1169.
\bibitem{BFO} D.G.~Barci, C.D.~Fosco and L.E.~Oxman,
Phys. Lett. {\bf B375} (1996) 267.
\bibitem{BLQ}  C.P.~Burgess, C.A.~L\"utken and F.~Quevedo,
Phys. Lett. {\bf B336} (1994) 18.
\bibitem{CRV} J.L.~Cort\'es, E.~Rivas and L.~Vel\'azquez,
Phys. Rev. {\bf D53} (1996) 5952.
\bibitem{mar} P.A. Marchetti in Common Trends in
Condensed Matter and High Enery Physics, Chia Laguna, Italy,
Sep 1995. hep-th/9511100 report.
\bibitem{BQ}C.P.~Burgess and F.~Quevedo,
 Phys. Lett. {\bf B329}(1994) 457.
\bibitem{DNS1} P.H.~Damgaard, H.B.~Nielsen and R.~Sollacher,
Nucl. Phys. {\bf B385} (1992) 227;
Phys. Lett.  {\bf B296} (1992) 132.
\bibitem{DNS5} P.H.~Damgaard, F.~De Jonghe and R.~Sollacher,
Nucl. Phys. {\bf B454} (1995) 701.
\bibitem{Jac} S.~Deser, R.~Jakiw and S.~Templeton,
{\it Phys. Rev. Lett.}
{\bf 48} (1982) 975; {\it Ann. of Physics (N.Y)} {\bf 140} (1982),
372.
\bibitem{Red} A.N.~Redlich, Phys. Rev. Lett. {\bf 52} (1984) 18;
Phys. Rev. {\bf D29} (1984)236.
\bibitem{GRS} R.~Gamboa Sarav\'\i ~,G.~Rossini and F.A.~Schaposnik, Int. J.
of Mod. Phys. {\bf A11} (1996) 2643.
\bibitem{VN} P.K.~Townsend, K.~Pilch and
P.~van~Nieuwenhuizen, Phys.~Lett. {\bf B136} (1984) 38; {\bf B137} (1984) 443.
\bibitem{DJ} S.~Deser and R.~Jackiw, Phys.Lett. {\bf B139} (1984) 371.
\bibitem{AFZ} I.J.R.~Aitchison, C.D.~Fosco and J.~Zuk,
Phys. Rev. {\bf D48} (1993) 5895.
\bibitem{Gross} D.Gross in {\it Methods in Field Theory}, Eds. R.
Balian and J. Zinn-Justinn, North-Holland, 1976.
\bibitem{witten} E.Witten, Comm. Math. Phys. {\bf 121} (1989) 351.
\bibitem{colemanlibro} See for example S.Coleman,
{\it Aspects of Symmetry}, Cambridge University Press, Cambridge, 1985.
\bibitem{BJ} D.G.~Boulware and R.~Jackiw, {\it Phys. Rev.} {\bf 186}
(1969), 1442.
\bibitem{Ch} M.S.~Chanowitz, {\it Phys. Rev.} {\bf D 2}
(1970) 3016; {\bf D 4} (1971), 1717.
\bibitem{dVG} H.J.~de Vega and H.O.~Girotti, {\it Nucl. Phys.}
{\bf B 79}
(1974), 77.
\bibitem{BJL} J.D.~Bjorken, {\it Phys. Rev.} {\bf 148}
(1966) 1467.
\bibitem{BJL2} K.~Johnson and F.E.~Low, {\it Prog. Theoret. Phys.
(Kyoto)},
Suppl. {\bf 37-38} (1966), 74.
\bibitem{Jac2} For a review see. e.g.
R.Jackiw {\it in} {`` Lectures on Current
Alebra and its Applications''}, (eds. S.B.~Treiman, R.~Jackiw and
D.J.~Gross), Princeton University Press, 1972.
%
\bibitem{Ban} R.~Banerjee, {\it Phys. Lett.} {\bf B 358} (1995), 297.
%
\bibitem{bralic} N.~Brali\'c, Phys.~Rev. {\bf D22} (1980) 3090.
\bibitem{jaco}K. Fujikawa, Phys. Rev. Lett. {\bf 42} (1979) 1195;\\
Phys. Rev. D{\bf 21} (1980) 2848; erratum-{\em ibid.\/} D{\bf 22}
(1980) 1499;\\
Phys. Rev. D{\bf 29} (1984) 285.
\end{thebibliography}
\end{document}